\documentclass[]{imag-ms-template}

\usepackage{soul} 
\usepackage{xcolor}
\usepackage{lineno}
\usepackage{multirow}

\title{Tau-BNO: Brain Neural Operator for Tau Transport Model} 
\author{
Nuutti Barron$^{1,*}$, Heng Rao$^{2,*}$, Urmi Saha$^{4}$, Yu Gu$^{2}$, Zhenghao Liu$^{2}$, Ge Yu$^{2}$,\\
Defu Yang$^{3}$,  Ashish Raj$^{1,\dagger}$, Minghan Chen$^{4,\dagger}$\\[6pt]
{\small $^{1}$Department of Radiology and Biomedical Imaging, University of California San Francisco, USA}\\
{\small $^{2}$College of Computer Science and Engineering, Northeastern University, China}\\
{\small $^{3}$Department of Computer Science, Hangzhou Normal University, China}\\
{\small $^{4}$Department of Computer Science, Wake Forest University, USA}\\
{\small $^{*}$Equal contribution;}
{\small $^{\dagger}$Equal co-advising}\\ 
{\small Corresponding authors: \texttt{chenm@wfu.edu}, \texttt{ashish.raj@ucsf.edu}}
}

\date{}
\begin{document} 

\maketitle 

% \linenumbers

\begin{abstract}

Mechanistic modeling provides a biophysically grounded framework for studying the spread of pathological tau protein in tauopathies such as Alzheimer's disease and frontotemporal dementia. 
Existing approaches typically model tau propagation as a diffusive process on the brain's structural connectome, reproducing macroscopic spatiotemporal patterns but neglecting the cellular transport and reaction mechanisms that govern microscale dynamics.
The Network Transport Model (NTM) was introduced to fill this gap, explaining how region-level progression of tau emerges from microscale biophysical processes. 
However, the NTM faces a common challenge for complex models defined by large systems of partial differential equations: the inability to perform parameter inference and mechanistic discovery due to high computational burden and slow model simulations. 
To overcome this barrier, we propose Tau-BNO, a Brain Neural Operator surrogate framework for rapidly approximating NTM dynamics that captures both intra-regional reaction kinetics and inter-regional network transport.
Tau-BNO combines a function operator that encodes kinetic parameters with a query operator that preserves initial state information, while approximating anisotropic transport through a spectral kernel that retains directionality. 
Empirical evaluations demonstrate high predictive accuracy ($R^2\approx$ 0.98) across diverse biophysical regimes and an 89\% performance improvement over state-of-the-art sequence models like Transformers and Mamba, which lack inherent structural priors. By reducing simulation time from hours to seconds, we show that the surrogate model is capable of producing new insights and generating new hypotheses. This framework is readily extensible to a broader class of connectome-based biophysical models, showcasing the transformative value of deep learning surrogates to accelerate analysis of large-scale, computationally intensive dynamical systems.

% Model inference allows the NTM to be used as a tool for biological discovery from empirical tau data, motivating the need to speed up simulation of the NTM to make inference tractable.

% However, conventional neural operators often conflate initial conditions with system coefficients, limiting generalization in directed networks. 

\end{abstract}

\section{Introduction}

% We introduce the Brain Neural Operator (BNO), a general neural operator framework for modeling molecular and connectomic dynamics in the brain. In this study, we instantiate the framework as BNO-Tau, specialized for modeling the spatiotemporal transport of tau across the mouse connectome.

% Rao: Literature
% Motivation (Rao)

The accumulation and spread of pathologically toxic tau protein in the brain is a hallmark of neurodegenerative tauopathies such as Alzheimer's disease (AD) and Frontotemporal Dementia (FTD), characterized by an initial spatially sparse accumulation, considered the seed, followed by a stereotyped spread to other regions over time \cite{Braak1991, Braak2006d}. 
Computational modeling has previously shown that tau spread is mediated by the long range neuronal white matter tracts interconnecting brain regions \cite{raj2012network, Wu2013, Katsikoudi2020}. 
Attempts to capture these spread dynamics generally model tau spread as a simple diffusive process on the brain's structural connectome (SC), a graph structure that represents brain regions as nodes, and white matter tract connections as weighted edges \cite{raj2012network, Fornari2019, Iturria-Medina2013, Schafer2020, Weickenmeier2018, Weickenmeier2019, Bertsch2023}. 
These connectome-based diffusion models have proven effective in staging diseases such as AD, predicting disease progression, and have shown promise as tools to uncover genes, cell types, and other biomarkers linked to tau spread in the brain \cite{torok2025searching, anand2022effects, vogel2021four, anand2025selective}. 

%\cite{anand2022effects, anand2025selective}

However, network diffusion models fail to capture the microscopic biophysical cellular mechanisms by which tau protein is transported in the brain, such as active axonal transport via dynein and kinesin motor proteins \cite{cuchillo2008phosphorylation, rodriguez2013tau, stern2017phosphoregulation, dixit2008differential, chaudhary2018tau}. Importantly, the inclusion of transport processes is key to understanding the divergent spatial patterning of tau under different seeding and conformational conditions \citep{torok2021emergence, torok2025directionality}---an aspect that has remained inscrutable by previous mathematical models. The Network Transport Model (NTM) \cite{tora2025network, barron2026biophysically} was recently proposed to capture both diffusive spread and microscopic transport mediated spread of tau in the brain, linking the biophysical cellular mechanisms mediating tau transport to observed spatiotemporal patterns of tau spread at the whole brain level. The mathematical description of NTM is provided in section \ref{sec:NTM}. The NTM and related models are an active area of study and represent the next stage of advancement in mechanistic modeling of dementias. 

Unfortunately, these emerging models pose a formidable computational challenge due to large-scale coupled partial differential equations (PDEs). Traditional finite element numerical solvers \citep{zienkiewicz2005finite} require solving PDEs individually for each parameter configuration, resulting in prohibitively high computational costs when exploring large parameter spaces. A quasi-static approximation to NTM was proposed to achieve numerical feasibility \cite{barron2026biophysically}. On a structural connectivity graph of size $n\times n$, it iteratively solves an edge-wise problem involving $o(n^2)$ numerical integration via MATLAB's \textit{ode45}, followed by a boundary value shooting problem at each node involving $o(n)$ calls to a nonlinear solver, i.e., MATLAB's \textit{fsolve} function (see \cite{barron2026biophysically} for details). 
The process repeats for each time point. Although this approach represents a meaningful advancement, it remains computationally demanding and extremely slow. 
Even with parallelized edge computations, a single 12-month NTM simulation on a 426-region brain parcellation takes approximately 10 hours. The slow simulation speed prohibits parameter inference of the NTM on empirical data, which can require tens of thousands of forward simulations. Such computational expense renders large-scale parameter inference, individualized model fitting, and clinical translation practically infeasible, motivating the need for a rapid and scalable surrogate model.

% This study aims to explore deep learning-based surrogate models to rapidly generate accurate NTM simulations under diverse pathological features, seeding sites, and biophysical parameter regimes. 
Unlike conventional finite-dimensional prediction tasks in computer vision \citep{he2016deep,krizhevsky2012imagenet}  or natural language processing \citep{devlin2019bert,brown2020language}, constructing a surrogate model for NTM requires approximating a solution operator that maps initial tau concentration fields and kinetic parameters to spatiotemporal trajectories \citep{kovachki2023neural}. Each configuration of system parameters defines a distinct PDE-governed dynamical system corresponding to a unique pathophysiological scenario. Consequently, a robust surrogate entails learning operators over infinite-dimensional function spaces rather than finite-dimensional vector mappings typical of conventional prediction tasks.
Recent advances in operator learning provide a promising proxy for accelerating scientific simulation \citep{kovachki2023neural,azizzadenesheli2024neural,liu2025architectures}.
By learning solution operators, these methods enable efficient approximation of PDE dynamics across varying initial conditions and parameter regimes without explicit numerical solvers at inference time. Architectures such as DeepONet \citep{lu2021learning}, Fourier Neural Operator (FNO) \citep{li2020fourier}, Wavelet Neural Operator (WNO) \citep{tripura2023wavelet, gupta2021multiwavelet}, Graph Neural Operator (GNO) \citep{li2020neural}, and the Geometry-Informed Neural Operator (GINO) \citep{li2023geometry}, have demonstrated success in accelerating simulations of complex physical systems with intricate geometries, as detailed in Discussion~\ref{sec:discussion}. 

% Moved related work to Discussion

% Limitations of Existing Neural Operators for Tau Transport Modeling
Despite this progress, existing neural operator frameworks are not directly suited to whole-brain tau transport modeling.
First, tau propagation involves globally coupled dynamics across $n$ brain regions and $o(n^2)$ structural connections over long temporal horizons, producing a high-dimensional function space with structured network dependencies. Most neural operators are designed for regular Euclidean grids and struggle to capture such connectome-mediated dynamics.
Second, many architectures process initial conditions and system parameters jointly within a single operator network. This entanglement limits interpretability and reduces generalization across heterogeneous parameter regimes, which is essential for individualized disease modeling and parameter inference.
Third, graph-based operator designs often rely on symmetric nearest-neighbor message passing without explicitly incorporating biologically meaningful edge attributes, such as connection strength or transport direction. Consequently, they cannot faithfully represent the intrinsic directionality of tau transport along axonal projections nor enforce anatomically grounded propagation along empirical white matter pathways \citep{tong2020directed, zhang2021magnet, tora2025network, raj2012network, torok2021emergence, franzmeier2019functional}.

% Despite these advances, existing neural operator frameworks face critical challenges when applied to our specific context---whole brain tau protein propagation modeling.
%First, tau transport involves intricate interactions among $n$ brain regions via their $o(n^2)$ connections, across hundreds of temporal steps, creating a high dimensional function space with complex global dependencies that span spatial networks and temporal evolution. Traditional neural operators, designed for problems on regular Euclidean grids, struggle to efficiently capture such globally-coupled, network-mediated dynamics.
% Second, existing methods typically encode initial conditions and system coefficients into unified representations, conflating physically distinct quantities with fundamentally different mechanistic roles. 
% This mixed encoding obscures individual contributions and diminishes generalization capacity across diverse coefficient regimes, which is critical for capturing individual-specific disease trajectories.
% Third, standard graph neural operators assume undirected topologies and cannot accurately represent the inherent directionality of tau spreading along axonal projections \citep{tong2020directed,zhang2021magnet}. 
% Fourth, they do not naturally incorporate known anatomical connectivity as explicit inductive bias, fundamentally limiting their ability to capture biologically realistic propagation dynamics on the connectome \citep{tora2025network,raj2012network,torok2021emergence,franzmeier2019functional}.

% Contribution

We introduce Tau-BNO, a specialized neural operator framework for spatiotemporal tau transport across the mouse connectome, which addresses these limitations through three key architectural innovations.
\textbf{1) Brain Operator Design}. Tau-BNO operates natively on irregular structural connectivity graphs, enabling operator learning directly over large-scale brain networks rather than regular Euclidean grids. The architecture preserves global inter-regional dependencies while maintaining computational scalability across high-dimensional network domains.
\textbf{2) Decoupled State-Dynamics Representation}.
We introduce a structured factorization of the operator mapping that distinguishes initial state encoding from parameter-driven dynamical evolution. A Query Operator captures regional initial tau concentrations, while a Function Operator models the influence of system parameters on regional kinetics and transport dynamics. This design enhances interpretability and generalization across heterogeneous conditions and diverse parameter regimes.
\textbf{3) Connectome-Constrained Anisotropic Propagation.} Tau-BNO embeds empirical structural connectivity as a biological prior and models asymmetric transport along axonal projections. We develop a directed, connectome-constrained graph operator that incorporates weighted and directional edge information, enabling realistic representations of anisotropic and directionally biased tau propagation dynamics.

% % Contribution: concise
% To address these limitations, we introduce Tau-BNO, a biologically grounded neural operator framework explicitly designed for spatiotemporal transport in the connectome. 
% First, unlike standard operators that entangle initial protein concentrations and mechanistic coefficients, Tau-BNO employs a disentangled encoding architecture, comprising distinct Query and Function Operators to independently model these two critical components.
% Second, to capture the inherent asymmetry of brain connectivity, we introduce a Directed Graph Operator that approximates anisotropic message passing based on the empirical structural connectome, overcoming the isotropic limitations of conventional graph convolutions. 
% Collectively, by integrating these biophysical and structural priors directly into the learning process, Tau-BNO enables spatiotemporally accurate prediction of pathological spreading across the mouse brain.

% Comprehensive ablation studies confirm that all three architectural components, including the Query Operator, Function Operator, and Directed Graph Operator, are essential for accurate spatiotemporal prediction of tau transport.
Comprehensive ablation studies demonstrate that each architectural component is necessary for accurate spatiotemporal prediction of tau transport dynamics. Tau-BNO achieves high predictive fidelity $R^2\approx 0.98$ value and reduces error by 89\% over large sequence-modeling approaches such as transformer and mamba, highlighting its superior capacity to model biologically grounded tau propagation. 

By providing rapid and high-fidelity approximations of the NTM, Tau-BNO functions as an efficient surrogate simulator that eliminates the computational bottlenecks associated with the previous quasi-static formulation. This acceleration renders large-scale parameter inference tractable, enabling systematic fitting of mechanistic NTM parameters to empirical tau imaging data. This will allow researchers to rapidly fit similar models to empirical data to extract patient-specific system coefficients, evaluate heterogeneous disease trajectories, and explore how specific cellular mechanisms could be targeted by future therapeutics. Beyond predictive accuracy, Tau-BNO enables mechanistic interrogation of tau transport processes. The framework supports inference of transport directionality, including anterograde versus retrograde bias, aggregation and fragmentation balance, cellular uptake and release rates, and spatial seeding patterns. These inferred parameters can be compared across distinct tau strains, linking molecular features such as structural conformations and phosphorylation states to macroscopic propagation dynamics. Moreover, fitted NTM parameters provide a principled basis for identifying associated biomarkers, including relevant cell types and gene expression signatures. 

 The surrogate model is capable of producing new insights and generating new hypotheses. The proposed Brain Neural Operator framework is a scalable computational foundation for investigating broader neuroscientific processes, such as axonal transport dynamics, and readily extensible to a broader class of connectome-based biophysical models of neurodegeneration. We anticipate that operator-based surrogate modeling will become a foundational component of next-generation computational platforms for mechanistic neuroscience.

% Tau-BNO also provides insights regarding the biophysical mechanisms driving tau spread in the brain, such as whether tau transport occurs with an anterograde or retrograde bias in neurons, where the balance between tau aggregation and fragmentation lies, how cellular tau uptake and release rates compare, and what inferred tau seeding looks like. 
% These can be compared across different strains of tau protein, linking specific tau protein structural elements and phosphorylation patterns to biophysical mechanisms of spread. 
% A fitted NTM can also be used to identify biomarkers, such as cell types and genes, that are linked to tau spread. Although the current study focuses on a specific tau transport model, its learnings can be readily extended to broader phenomena in neurodegeneration. We anticipate that the BNO framework will become a critical component of emerging computational platforms in neurology.

 % However, because the NTM provides a link to the underlying biophysical mechanisms driving tau spread in the brain SC network, the NTM can go one step further and link these biomarkers to specific cellular mechanisms mediating tau spread, which diffusion based models are unable to accomplish.
\section{Results}

\begin{figure}[htbp]
\centering
\includegraphics[width=0.95\textwidth]{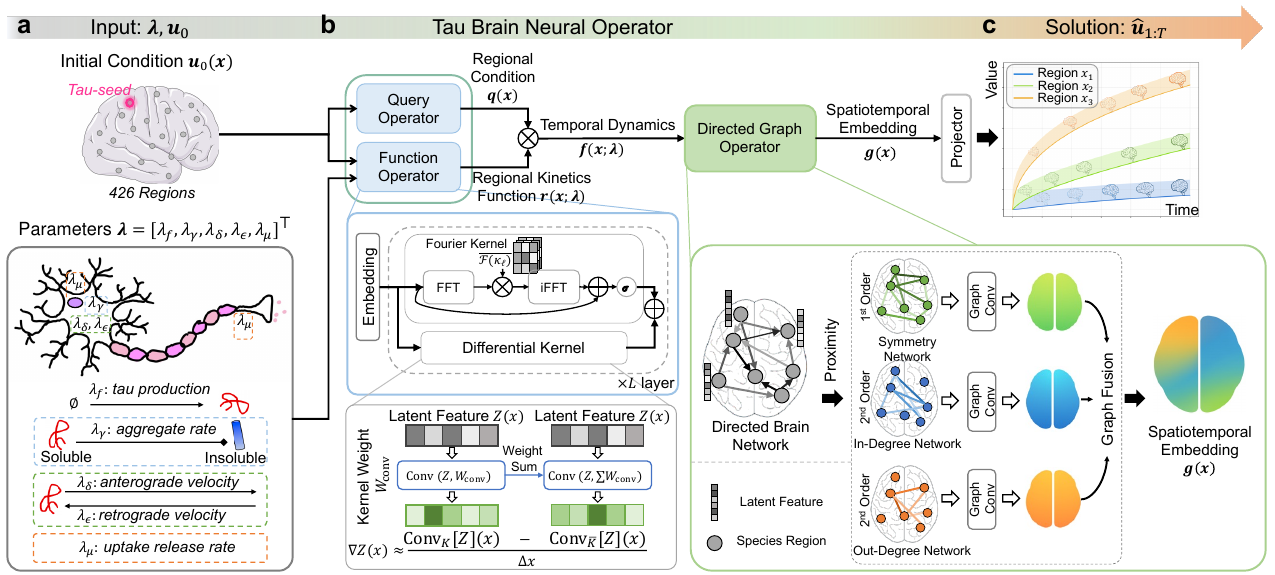}
\caption{\textbf{Tau-Brain Neural Operator (Tau-BNO) Architecture.}
\textbf{a.}  Input consists of initial tau concentration fields derived from experimentally defined mouse injection sites, together with sampled aggregation and fragmentation rate parameters  ($\lambda_f, \lambda_\gamma, \lambda_\delta, \lambda_\epsilon, \lambda_\mu$).
\textbf{b.}  A Query Operator encodes regional initial tau concentrations, and a Function Operator encodes kinetic parameters; their interaction is passed through a Directed Graph Operator defined on the structural connectome to model asymmetric inter-regional transport.
\textbf{c.}  A learnable projector generates tau spatiotemporal dynamics across all 426 brain regions.}
\label{fig:Tau-BNO}
\end{figure}

\begin{figure}[htbp]
\centering
\includegraphics[width=0.8\textwidth]{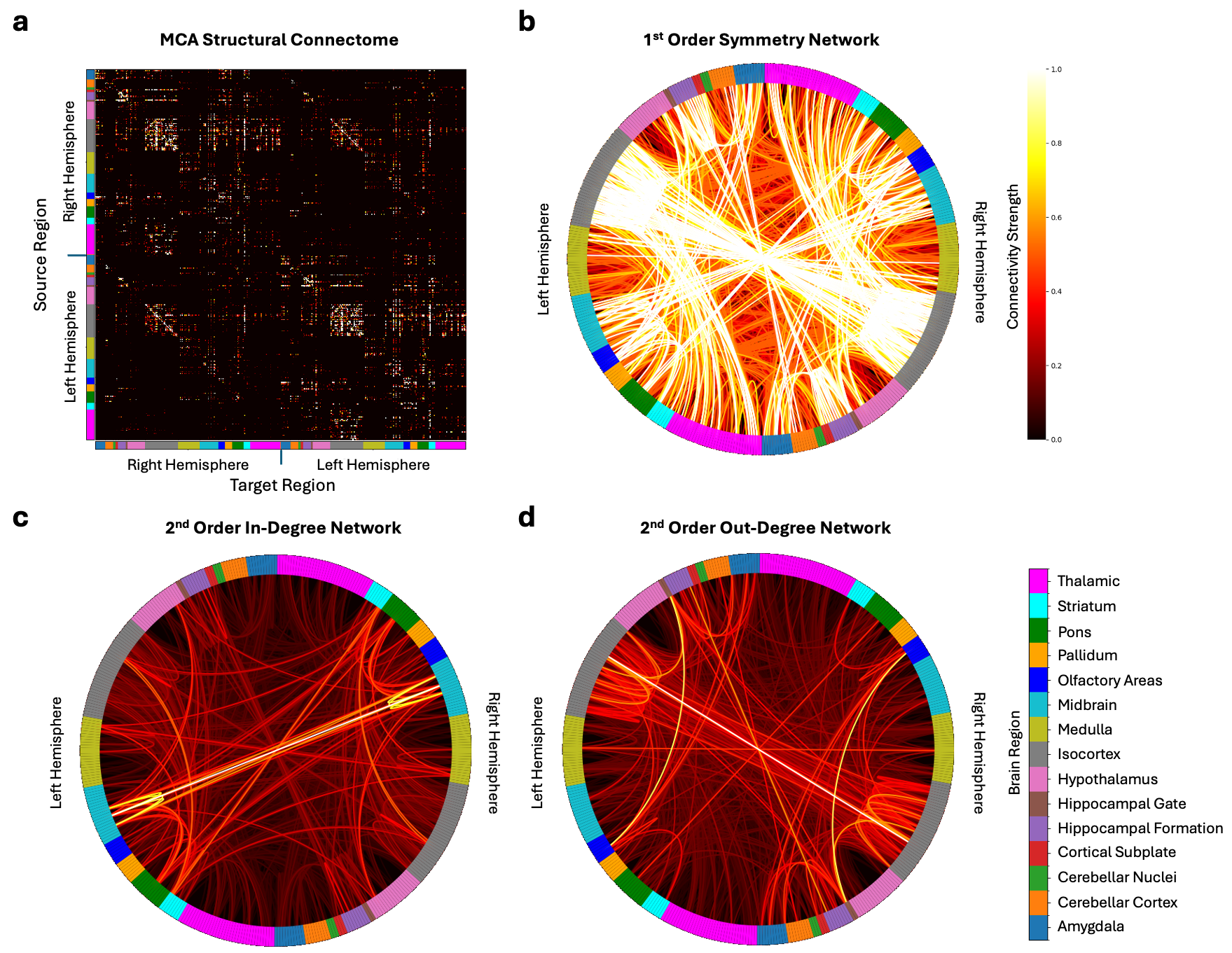}
\caption{\textbf{Brain network representations for tau transport modeling.} 
\textbf{a.} Original directed MCA connectome encoding region-to-region tau transport pathways. 
\textbf{b--d.} Three derived undirected proximity networks capturing complementary topological features: 
\textbf{b.} first-order symmetric network preserving global connectivity structure; 
\textbf{c.} second-order indegree network emphasizing convergent (afferent) connectivity; 
\textbf{d.} second-order outdegree network emphasizing divergent (efferent) connectivity.
Edge brightness indicates connection strength. Together, these representations enable graph neural networks to recover directional transport dynamics from symmetrized adjacency matrices.}
\label{fig:BrainNetwork}
\end{figure}

We develop Tau-BNO to rapidly generate accurate simulations of the Network Transport Model (NTM) across diverse seeding sites and biophysical parameter regimes. The Tau-BNO architecture (Figure~\ref{fig:Tau-BNO}) comprises two complementary components: a Query Operator (QO) for encoding regional initial tau concentrations and a Function Operator (FO) for representing kinetic parameters governing transport dynamics. These latent representations are subsequently integrated through a Directed Graph Operator (DGO), which projects the learned dynamics onto the empirical mouse structural connectome comprising 426 anatomical regions derived from the mouse mesoscale connectivity atlas (MCA) produced by the Allen Institute \cite{oh2014mesoscale}.
To enable efficient graph convolution operations \citep{kipf2016semi} while preserving the directional structure inherent in tau transport, we construct three network representations derived from the original directed connectome (Figure~\ref{fig:BrainNetwork}). By jointly leveraging these structural priors, the model reconstructs asymmetric transport dynamics and implicitly learns a data-driven directed network that reflects physiologically grounded axonal pathways, ensuring accurate modeling of directional tau propagation.

% our method
% We propose Tau-BNO (Tau Brain Network Operator) as shown in Fig.~\ref{fig:Tau-BNO}, a neural network architecture designed to replace traditional PDE solvers while achieving superior inference speed and accuracy compared to both classical numerical methods and state-of-the-art neural operators.

% %% Our contribution (Model Architecture)
% Our model addresses a key limitation of existing neural operators: the conflation of initial conditions and system coefficients into a unified representation. 
% Tau-BNO explicitly decouples these two sources of information through parallel processing pathways, enabling independent learning of regional condition features and regional kinetic functions. These pathways are integrated through a Directed Graph Operator (DGO) that leverages known anatomical brain connectivity to capture region to region interactions across the 426 region parcellation, naturally incorporating the directionality and heterogeneity of tau propagation pathways.
% The Tau-BNO architecture (Figure~\ref{fig:Tau-BNO}) is constructed to explicitly operationalize the separation of biological state from kinetics. 
% Instead of a unified representation, the model employs parallel processing pathways that independently encode regional initial conditions and kinetic coefficients. These distinct features are then integrated via a Directed Graph Operator (DGO), which maps the learned latent representations onto the empirical mouse connectome (consisting of 426 anatomical regions).

\subsection{Baseline Methods}
We benchmarked Tau-BNO against 11 diverse methods spanning six architectural paradigms and five neural operator variants. 
The generated dataset was split into 80\% training, 10\% validation, and 10\% testing sets. 
Each model was tasked with mapping the initial conditions and system parameters $\{\mathbf{u}_0, \boldsymbol{\lambda}\}$ to the spatiotemporal evolution $\{\mathbf{u}_{1:T}\}$ over $T$ time steps. 
To ensure fair comparison across architectures, the system parameters $\boldsymbol{\lambda}$ were broadcast to match the spatial dimensions of the state variables, providing consistent input representations for all models.

Classical baselines included Multi-layer Perceptrons (MLP) \citep{rosenblatt1958perceptron}, Kolmogorov-Arnold Networks (KAN) \citep{liu2024kan}, and NeuralODE \citep{chen2018neural} for continuous-time modeling. 
We further evaluate state-of-the-art sequence modeling architectures, including Transformer \citep{vaswani2017attention} and Mamba \citep{gu2024mamba}, a recent state space model renowned for its strong performance in capturing long range dependencies.
Critically, we compared against a family of advanced neural operators designed to learn mappings between function spaces, including DeepONet \citep{lu2021learning} as a foundational operator learning architecture, the Fourier neural operator (FNO) \citep{li2020fourier} with spectral convolutions, multiwavelet-based operator models (MWT) \citep{gupta2021multiwavelet} and wavelet neural operators (WNO) \citep{tripura2023wavelet} for multi-scale modeling, GINO \citep{li2023geometry} for irregular geometries, and the localized neural operator (LNO) \citep{liu2024neural}, which incorporates differential kernels for fine-grained spatial resolution.

All models were evaluated using three complementary metrics: Root Mean Square Error (RMSE), Mean Absolute Error (MAE), and relative $\ell_2$ error for trajectory prediction. Metrics were computed across all 426 brain regions and all temporal sampling points in the held-out test set. For a single trajectory variable, the evaluation metrics are defined in Eq.~\ref{eq:evaluation_metrics}, where $T$ denotes the number of time steps.

% \begin{equation}
% \begin{aligned}
% \text{RMSE} &= \sqrt{\frac{1}{T} \sum_{i=1}^{T} (\hat{u}_i - u_i)^2} \\[0.3em]
% \text{MAE} &= \frac{1}{T} \sum_{i=1}^{T} |\hat{u}_i - u_i| \\[0.3em]
% \text{L2 rel. err.} &= \frac{\sqrt{\sum_{i=1}^{T} (\hat{u}_i - u_i)^2}}{\sqrt{\sum_{i=1}^{T} u_i^2}}
% \end{aligned}
% \label{eq:evaluation_metrics}
% \end{equation}

\begin{equation}
\text{RMSE} = \sqrt{\frac{1}{T} \sum_{i=1}^{T} (\widehat{\mathbf{u}}_i - \mathbf{u}_i)^2}, \quad
\text{MAE} = \frac{1}{T} \sum_{i=1}^{T} |\widehat{\mathbf{u}}_i - \mathbf{u}_i|, \quad
\text{Rel.~L2} = \frac{\sqrt{\sum_{i=1}^{T} (\widehat{\mathbf{u}}_i - \mathbf{u}_i)^2}}{\sqrt{\sum_{i=1}^{T} \mathbf{u}_i^2}}
\label{eq:evaluation_metrics}
\end{equation}

\subsection{Performance Evaluation}

We benchmarked Tau-BNO against six general architectural baselines and five neural operator variants to evaluate the role of operator learning in tau-transport prediction and to assess the performance gains achieved by our proposed architecture.
All experiments were repeated three times with different random seeds, and we report the mean performance with standard deviations across runs. Figure~\ref{fig:MainResults} summarizes the comparative results. 
Additional experiments are provided in~\nameref{sec:extra_results}.

\subsubsection{Comparison with General Architectures}

% Paragraph 1: The dominance of Tau-BNO and the hierarchy of Sequence Models
As illustrated in Figure~\ref{fig:MainResults}a, Tau-BNO demonstrates a decisive advantage over general-purpose architectures, achieving an RMSE of $6.922 \times 10^{-6}$, with an 89\% improvement over the strongest baseline, Mamba.
Among general architectures, sequence models (Mamba and Transformer) significantly outperform standard MLP-based approaches. Mamba further surpasses the Transformer, likely due to its structured state-space formulation, which more effectively captures long-range temporal dependencies in extended trajectories than attention-based models with fixed context windows.
% Paragraph 2: The failure mode of MLP-based and Point-wise architectures (KAN, NeuralODE, DeepONet)
In contrast, architectures lacking explicit spatiotemporal dynamics, including KAN, NeuralODE, and the standard MLP, exhibit significantly higher error rates. Although KAN leverages the Kolmogorov–Arnold representation theorem for functional approximation and NeuralODE models continuous dynamics, neither approach adequately captures the high-dimensional (426-region) connectome-mediated transport structure. Despite being designed for operator learning, DeepONet achieves only marginal improvement over the MLP baseline (7\%). Its reliance on MLP-based branch and trunk networks limits its ability to model complex inter-regional coupling.
% Paragraph 3: The Theoretical Synthesis (Why they all failed vs Tau-BNO)
Collectively, these findings reveal a fundamental limitation of general architectures that operate on finite-dimensional vector mappings rather than explicitly learning operators over structured function spaces. Tau-BNO's superior performance demonstrates that accurate modeling of large-scale biological transport requires explicit incorporation of spatial graph coupling to capture the geometric structure of the underlying PDE solution manifold.

\subsubsection{Comparison with Advanced Neural Operators}
\begin{figure}[htbp]
\centering
\includegraphics[width=0.95\textwidth]{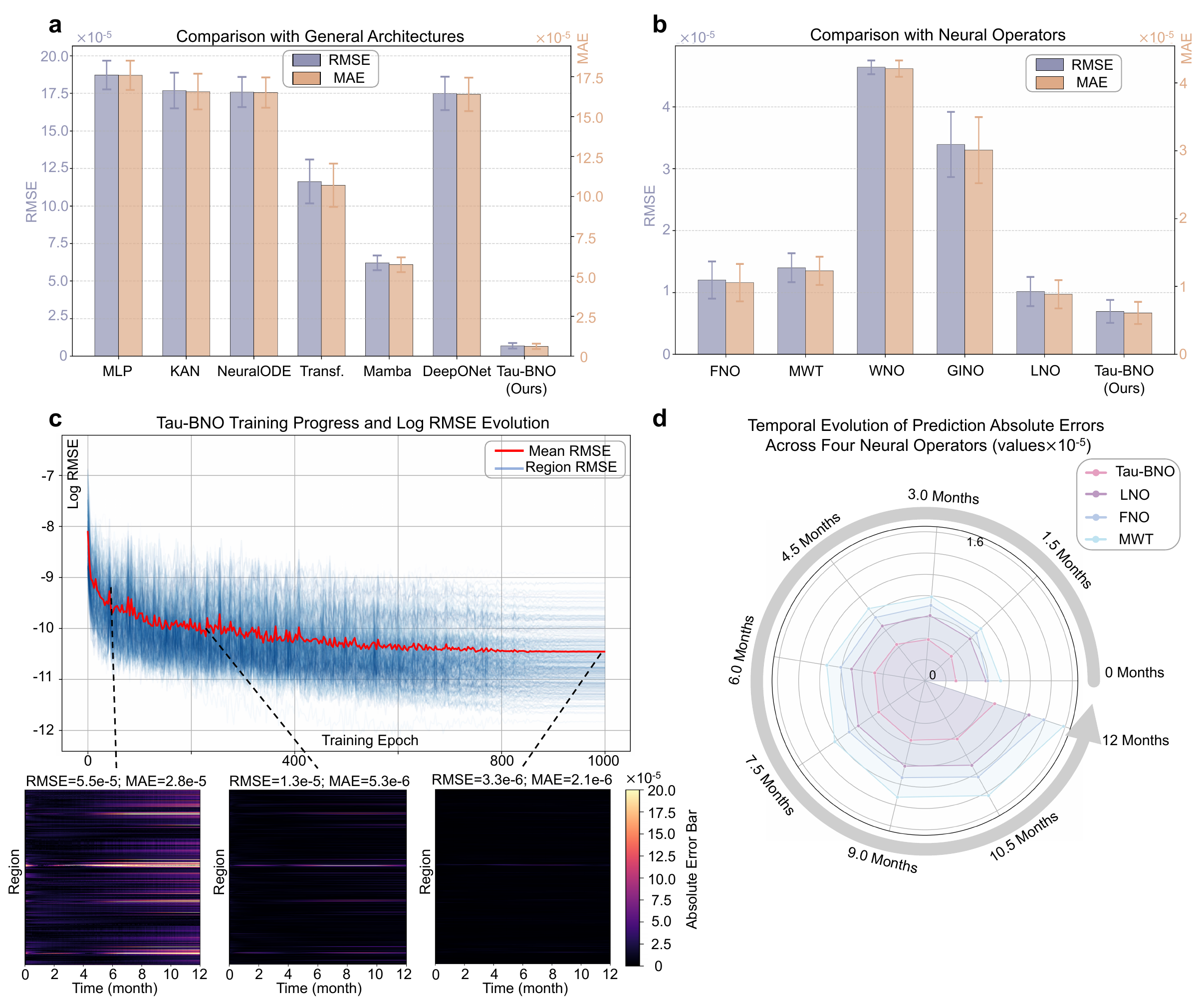}
\caption{\textbf{Performance comparison of Tau-BNO with 11 benchmarking models.} 
\textbf{a.} Quantitative evaluation (RMSE, MAE) across six general-purpose architectures, highlighting the limitations of standard predictive models for PDE-governed tau transport.
\textbf{b.} Comparison among five neural operator variants, showing Tau-BNO's superiority within the operator learning family. 
\textbf{c.} Training dynamics of Tau-BNO over 1000 epochs, showing log-scale RMSE and stable convergence behavior. 
\textbf{d.} Temporal evolution of absolute prediction error for Tau-BNO and three representative neural operators, illustrating sustained accuracy across the full trajectory.
}
\label{fig:MainResults}
\end{figure}

As illustrated in Figure~\ref{fig:MainResults}b, Tau-BNO establishes a new state-of-the-art, achieving an RMSE of $6.922 \times 10^{-6}$ and outperforming the leading neural operator baseline (LNO) by 32\%. 
This superior performance is consistent across temporal scales: Figure~\ref{fig:MainResults}d demonstrates that Tau-BNO maintains the lowest absolute error throughout the entire trajectory, indicating stable predictive accuracy over long horizons.
Operator architectures that emphasize localized representations show limitations in modeling the global transport dynamics of tau. Wavelet-based operators (WNO, MWT) underperform their Fourier-based counterparts, likely due to the locality of wavelet bases, which emphasize high-frequency interactions while attenuating the low-frequency spectral components that govern smooth, domain-wide diffusion. Similarly, GINO ($3.390 \times 10^{-5}$) is constrained by its reliance on local graph message passing, which limits its effective receptive field and impedes modeling of long-range dependencies across the connectome.
Although LNO improves upon FNO through localized differential kernels, it remains inferior to Tau-BNO. This gap indicates that incremental refinements of local operators are insufficient for capturing connectome-scale transport dynamics. By disentangling regional initial conditions from global kinetic parameters and embedding directional graph structure into the operator design, Tau-BNO provides a biologically grounded inductive bias that more effectively represents the underlying transport mechanism.

\subsection{Ablation Analysis of Tau-BNO Architecture}

To quantify the contribution of each architectural component in Tau-BNO, we performed systematic ablation studies targeting the Function Operator (FO), Query Operator (QO), and Directed Graph Operator (DGO), as well as the two kernel types used within the operator blocks. 
\subsubsection{Necessity of Modular Decomposition}

Figure~\ref{fig:Results_Ablation} quantifies the impact of ablating each architectural component. 
Eliminating the FO results in the most catastrophic performance degradation (RMSE = $1.203 \times 10^{-4}$), confirming that learning tau dynamics solely from initial conditions is insufficient. 
Explicit encoding of kinetic parameters is therefore necessary to capture the governing reaction–diffusion processes.
Removing the DGO produces a 29\% increase in RMSE ($8.93 \times 10^{-6}$), underscoring the importance of connectome topology for modeling inter-regional transport. Without graph propagation, predictions collapse toward spatially independent regional dynamics. Notably, this graph-ablated variant still outperforms the strongest baseline (LNO, RMSE = $1.014 \times 10^{-5}$) by 12\%, highlighting the strength of the decoupled operator design. 
%By encoding state and kinetic parameters, Tau-BNO retains meaningful dynamics even when topological priors are removed.
% DGO ablations
Ablating the Query Operator yields a modest 6\% increase in RMSE. This smaller performance degradation likely stems from the restricted range of initial condition configurations used during training. However, the QO remains essential for generalizing to heterogeneous initial conditions, such as complex disease stages where explicit disentanglement of state from dynamics is required to capture non-trivial propagation patterns.

\subsubsection{Complementarity of Fourier and Differential Kernels}

\begin{figure}[htbp]
    \centering
    \includegraphics[width=0.85\linewidth]{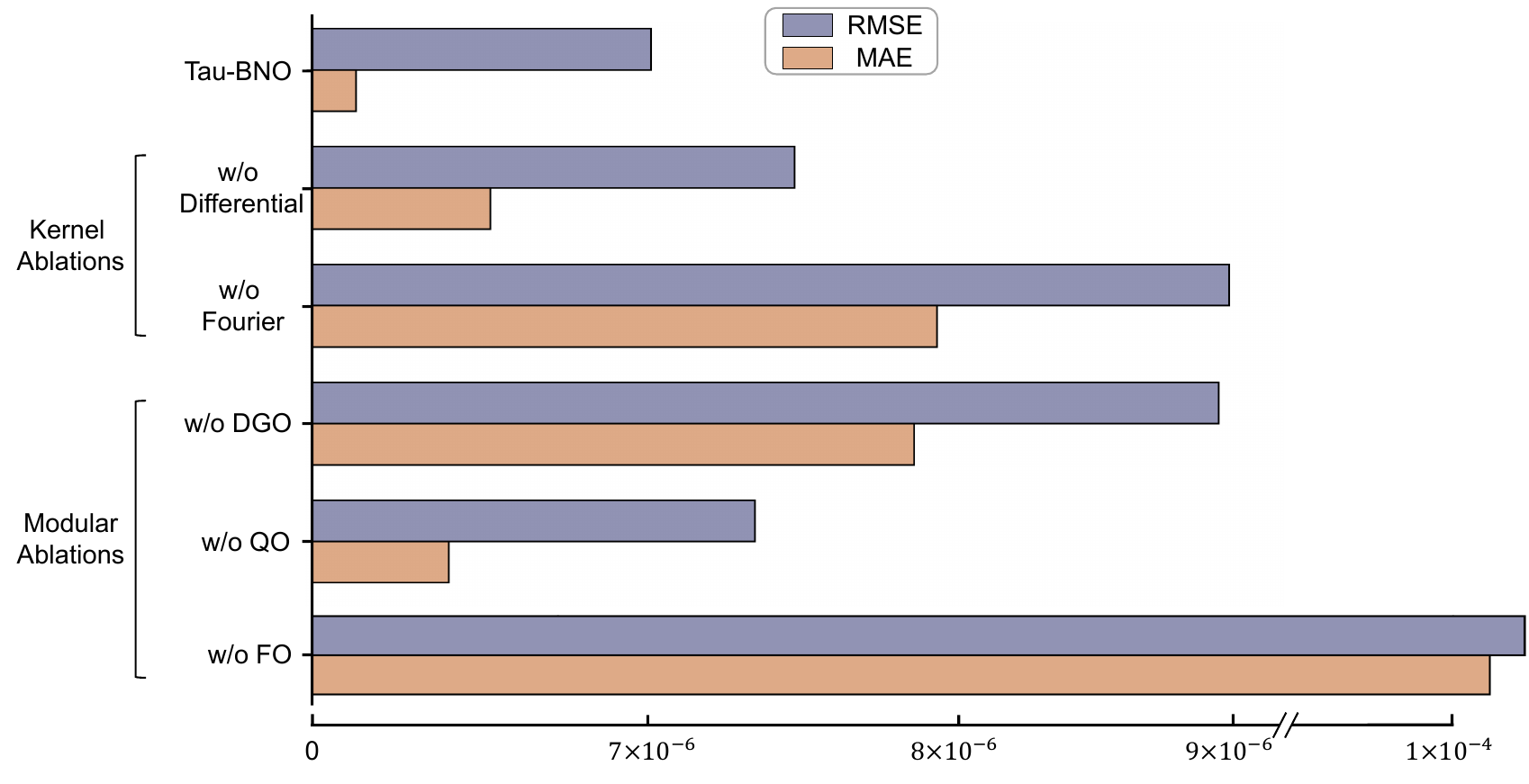}
    \caption{\textbf{Ablation analysis of Tau-BNO architecture.} Performance comparison for modular (FO, QO, DGO) and kernel (Fourier, differential) ablations.}
    \label{fig:Results_Ablation}
\end{figure}

To evaluate the role of multi-scale spatial representations, we ablated the Fourier and differential kernels in both operator blocks. As shown in Figure~\ref{fig:Results_Ablation}, both components contribute to performance, though with differing impact.
% main
Removing the Fourier kernel induces a substantial performance penalty (RMSE increased by 30\%). 
This finding indicates that spectral representations play a dominant role in modeling the spatially coherent, smooth variations of tau concentration fields, effectively capturing global dependencies across the continuous brain domain.
In contrast, ablating the differential kernel results in a moderate 8\% degradation. 
This performance asymmetry aligns with the biophysics of tauopathy: the dynamics are fundamentally governed by large-scale network diffusion (captured by global Fourier modes), with local reaction terms playing a secondary modulatory role (resolved by local differential operators).
% summary
These results support the architectural design of Tau-BNO: decoupling state and kinetic parameters, incorporation of connectome topology, and integration of multi-scale spatial operators are all necessary for accurate surrogate modeling of large-scale tau transport.

\subsection{Spatiotemporal Prediction Accuracy and Generalization}

\begin{figure}[htbp]
\centering
\includegraphics[width=1\textwidth]{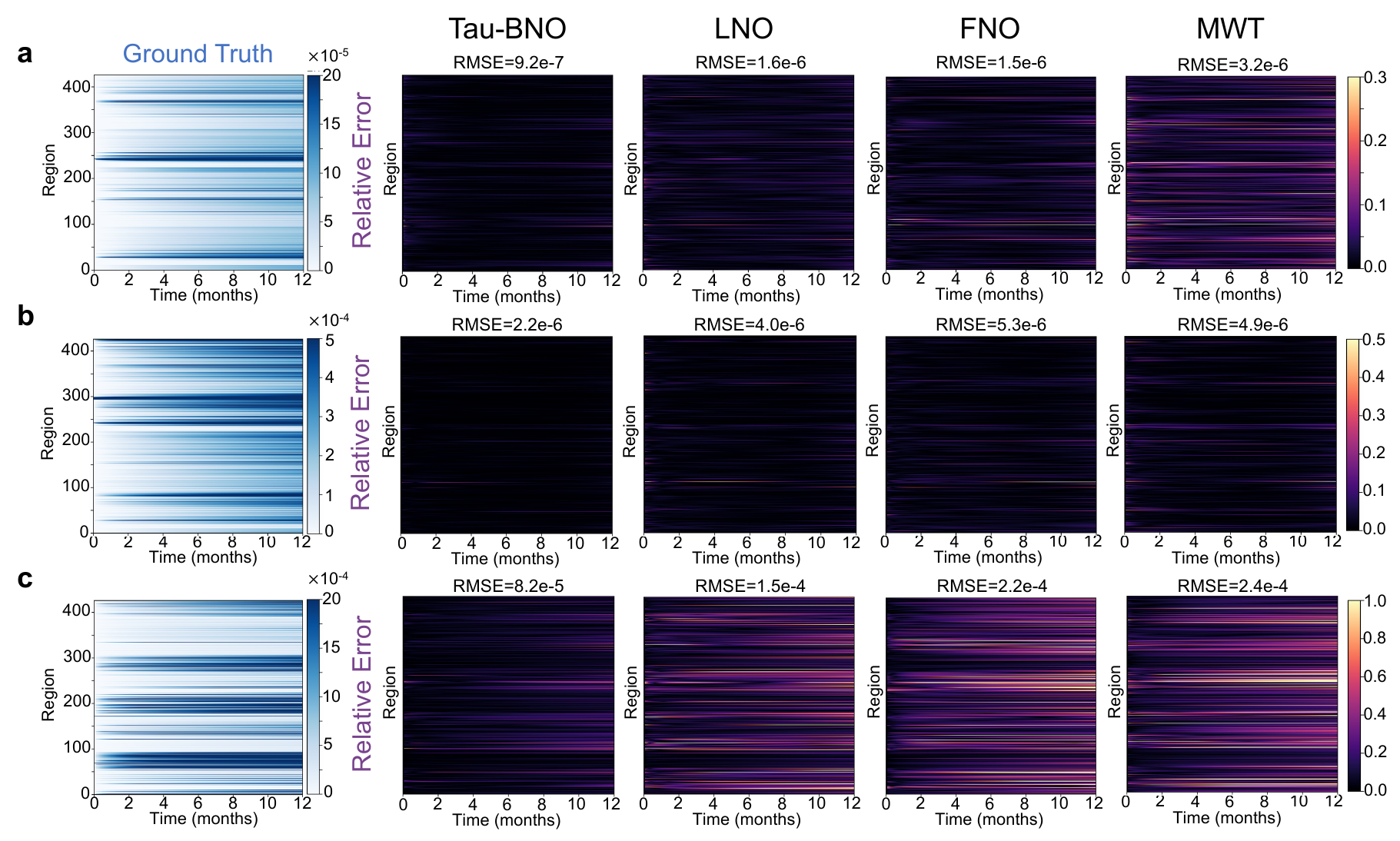}
\caption{\textbf{Spatiotemporal prediction accuracy across diverse physiological regimes.} 
Brain heatmaps of relative prediction error for Tau-BNO and three neural operator baselines under three representative biophysical regimes. 
Parameters $[\lambda_f, \lambda_\gamma, \lambda_\delta, \lambda_\epsilon, \lambda_\mu]$ denote tau production rate, aggregate rate, anterograde transport velocity, retrograde transport velocity, and uptake release rate, respectively.
\textbf{a.} Transport-only regime ($\lambda_f=0$): $[0.0, 8\times10^{-3}, 100, 100, 1.20]$.
\textbf{b.} High anterograde regime (anterograde dominant transport): $[9\times10^{-4}, 4.8\times10^{-3}, 80.6, 22.6, 0.47]$.
\textbf{c.} High retrograde regime (elevated production, strong retrograde transport): $[9.8\times10^{-3}, 4.2\times10^{-3}, 58.6, 92.5, 0.51]$.
Color scale denotes relative error magnitude. Across regimes, Tau-BNO exhibits consistently lower error than competing models.}
\label{fig:Results_heatmap}
\end{figure}

\subsubsection{Robust Generalization Across Diverse Physiological Regimes}
% Trajectory (differnet coeff), brain mapping, heatmap (differnet coeff)
% todo: Introducing heatmap comparison
% todo: Introducing Trajectory Prediction for few examples.

% \textbf{Superior regional prediction fidelity across NTM biophysical parameter space.}
Figure~\ref{fig:Results_heatmap} presents spatiotemporal error distributions for Tau-BNO and three leading neural operator methods across three representative biophysical parameter regimes. 
These regimes induce qualitatively distinct patterns of regional tau accumulation, ranging from rapid, widespread propagation (high anterograde regime) to slower, focal progression (transport-dominated regime), thereby providing a stringent test of generalization across diverse physiological conditions.
Tau-BNO consistently achieves the lowest error across the brain. In standard transport regimes (Figure~\ref{fig:Results_heatmap}a, b), mean relative error remains below 1\%. Under the high-retrograde stress condition (regime c), our method reduces error by 45–66\% relative to competing operator models, demonstrating robustness under physiologically extreme configurations.
% Tau-BNO consistently achieves substantially lower relative errors across all 426 brain regions compared to alternative architectures (mean relative L2 error: 0.84\% for regime a, 0.56\% for regime b, 4.8\% for regime c). 
% Notably, while all methods exhibit increased errors under the challenging high retrograde regime (c) characterized by high production and strong retrograde transport Tau-BNO's error remains 45--66\% lower than competing approaches, demonstrating superior robustness to extreme physiological conditions. 
% Moreover, competing methods show spatially heterogeneous failure modes with disproportionate errors, whereas Tau-BNO maintains spatially uniform accuracy.

% Furthermore, while baseline architectures exhibit spatially heterogeneous failure modes with disproportionate regional errors, Tau-BNO preserves uniform accuracy across all 426 brain regions.

\begin{figure}[htbp]
    \centering
    \includegraphics[width=1\linewidth]{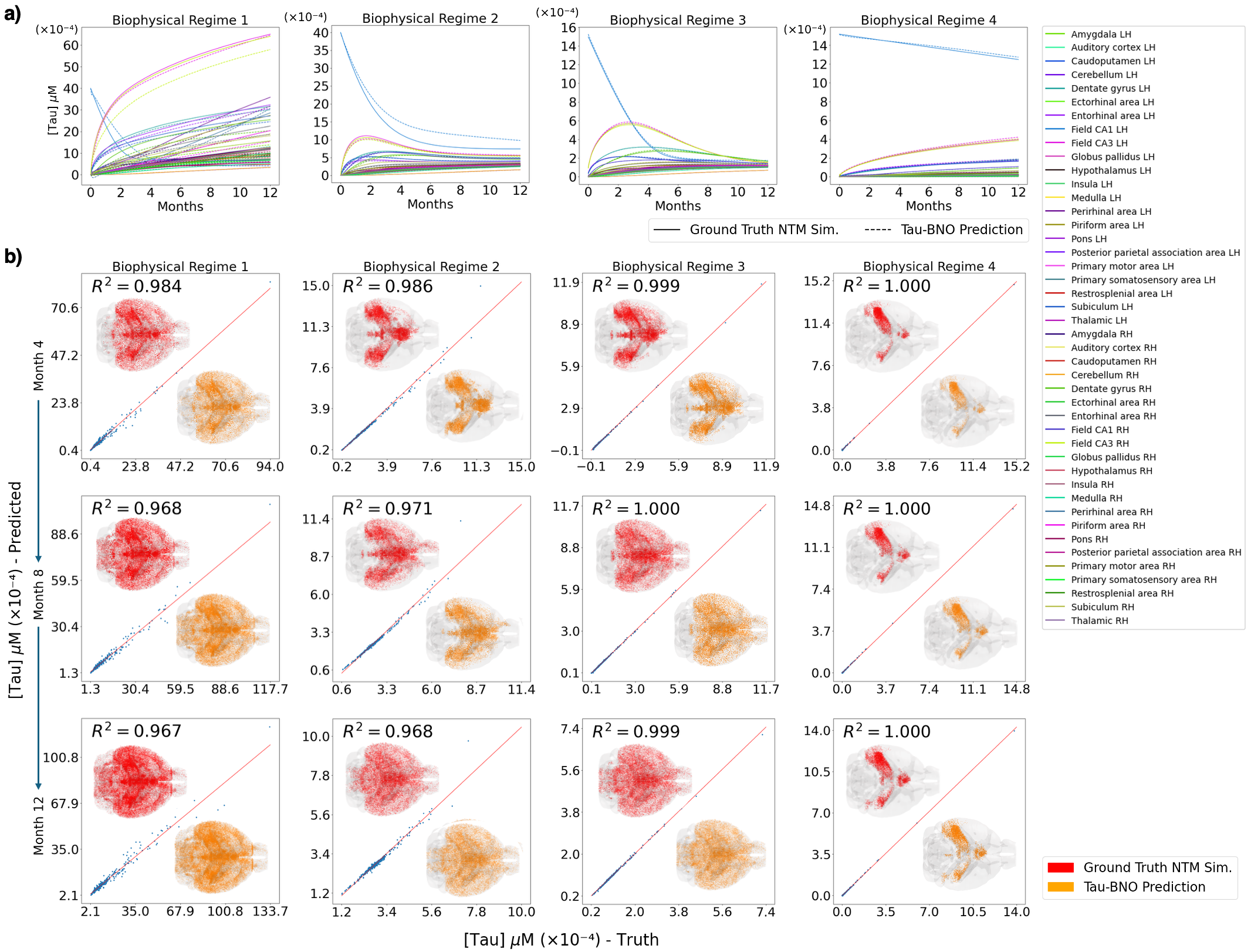}
    \caption{\textbf{Prediction accuracy under various and extreme biophysical parameter regimes. a.} Tau trajectories comparing ground truth NTM simulations (solid lines) and Tau-BNO prediction (dashed lines) for four biophysical parameter regimes, aggregated from the 426-region MCA atlas to the 44-region DS atlas. Regime specifications are provided in Supplementary Table~\ref{tab:bio-regimes}. \textbf{b.} Predicted versus ground truth spatial tau concentrations at 4-month intervals under four biophysical parameter regimes. Blue dots represent brain regions under the MCA mouse atlas. Insets show glass brain visualizations for ground truth (top left) and Tau-BNO (bottom right), with red/orange intensity reflecting regional tau concentration.}
    \label{fig:r2-fig}
\end{figure}

Tau-BNO demonstrates similarly high prediction accuracy in four additional representative NTM biophysical parameter regimes (Figure \ref{fig:r2-fig}a). These regimes were selected to probe boundary cases, including near-zero production rates, extreme transport asymmetries, elevated kinetic aggregation rates, as well as out-of-sample cellular uptake and release rates. The description of biophysical regimes is detailed in Supplementary Table~\ref{tab:bio-regimes}. These scenarios assess both interpolation within and extrapolation beyond the training distribution. Across regimes, predicted regional trajectories (dashed lines) closely follow ground truth NTM simulations (solid lines) across all time points, accurately capturing distinct dynamical behaviors including monotonic growth, transient peaks, and regional decline.
Additional evaluations across expanded kinetic parameter configurations are provided in Supplementary Results~\nameref{sec:Another_Trajectory_Prediction}.

% This robustness stems from explicitly decoupling system coefficients (Function Operator), initial conditions (Query Operator), and network topology (Directed Graph Operator), enabling the model to independently learn local production clearance kinetics, regional-specific initial tau distributions, and directional axonal transport processes each of which dominates under different coefficient regimes.

%To evaluate temporal prediction accuracy at finer granularity, we examined long term concentration trajectories in anatomically critical regions across six additional coefficient combinations spanning the explored coefficient space (Fig.~\ref{fig:Results_trajectory_sensitivity}). 

\subsubsection{Sustained Predictive Performance Across Timescales}
% R-square

Figure~\ref{fig:r2-fig} evaluates Tau-BNO's ability to reproduce spatial tau distributions across 426 brain regions over simulation horizons, compared against numerical PDE solutions under diverse NTM kinetic parameter configurations. Tau-BNO achieves remarkably high $R^2$ values between predicted and ground truth values, ranging from 0.967-1.000 in the selected model parameter configurations, even at later time points. Although predictive accuracy decreases gradually over time, the degradation remains limited. In the most challenging configuration, $R^2$ declines by 0.018 between months 4 and 12 (from 0.986 to 0.968). In the best-performing cases, $R^2$ remains at 1.000 across the evaluated time span, indicating sustained long-horizon stability.
The accompanying glass brain visualizations further corroborate these quantitative findings. Across time points and parameter regimes, Tau-BNO predictions (orange) closely replicate the spatial distribution of tau accumulation observed in the numerical PDE solutions (red), including both dominant high-concentration hubs and more diffuse inter-regional spread. Importantly, regional asymmetries and interhemispheric propagation patterns are preserved over time, demonstrating that predictive fidelity extends beyond aggregate metrics to anatomically coherent spatial structure.

\subsubsection{High Fidelity Prediction Across Varying Initial Tau Seeding.}
\begin{figure}[htbp]
    \centering
    \includegraphics[width=1\linewidth]{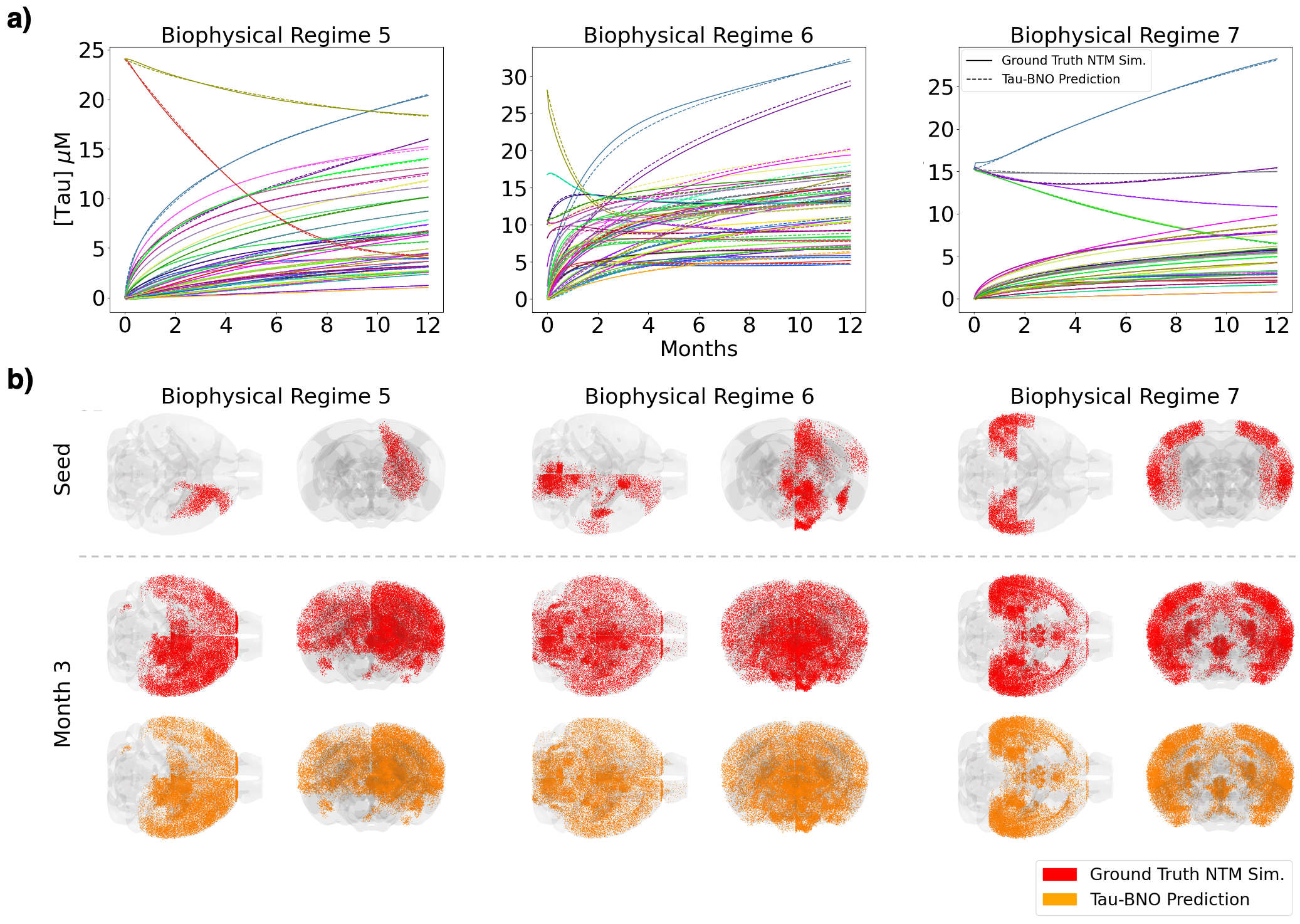}
    \caption{\textbf{Prediction accuracy across various initial seeding conditions. a.} Ground truth and Tau-BNO predicted tau trajectories under different seeding configurations, aggregated from the 426-region MCA atlas to the 44-region DS atlas. Region color legend for tau trajectories follows Figures \ref{fig:r2-fig}. \textbf{b.} Glass brain visualizations showing the seeded regions (top) and resulting spatial tau distributions 3 months post-seeding (bottom) across three representative NTM parameter regimes.}
    \label{fig:seeding}
\end{figure}

Figure~\ref{fig:seeding} demonstrates Tau-BNO's ability to accurately predict NTM solutions under diverse initial tau seeding configurations across biophysical regimes 5–7. In these experiments, kinetic parameters are randomly sampled while seeding locations, magnitudes, and the number of affected regions are systematically varied.
Figure~\ref{fig:seeding}a shows that the model adapts to these heterogeneous initializations, reproducing distinct temporal signatures induced by multi-focal versus localized seeding.
Figure~\ref{fig:seeding}b confirms spatial fidelity: glass brain visualizations show that Tau-BNO preserves seed localization and subsequent inter-regional propagation patterns, including hemispheric spread and large-scale diffusion structure. Even under rapid and widespread regimes, spatial distributions remain anatomically consistent with the numerical solutions.
Together, these results demonstrate robust generalization to heterogeneous initial conditions while maintaining spatiotemporal accuracy.
Collectively, these experiments establish Tau-BNO as a reliable and computationally efficient surrogate for NTM across a broad physiological parameter space and initial condition. By maintaining accuracy under extreme parameter regimes, long temporal horizons, and heterogeneous seeding conditions, Tau-BNO enables scalable inverse modeling, uncertainty quantification, and parameter inference that would otherwise be computationally prohibitive with traditional PDE solvers.

\subsection{Biophysical Exploration with Tau-BNO}

\begin{figure}[htbp]
    \centering
    \includegraphics[width=1\linewidth]{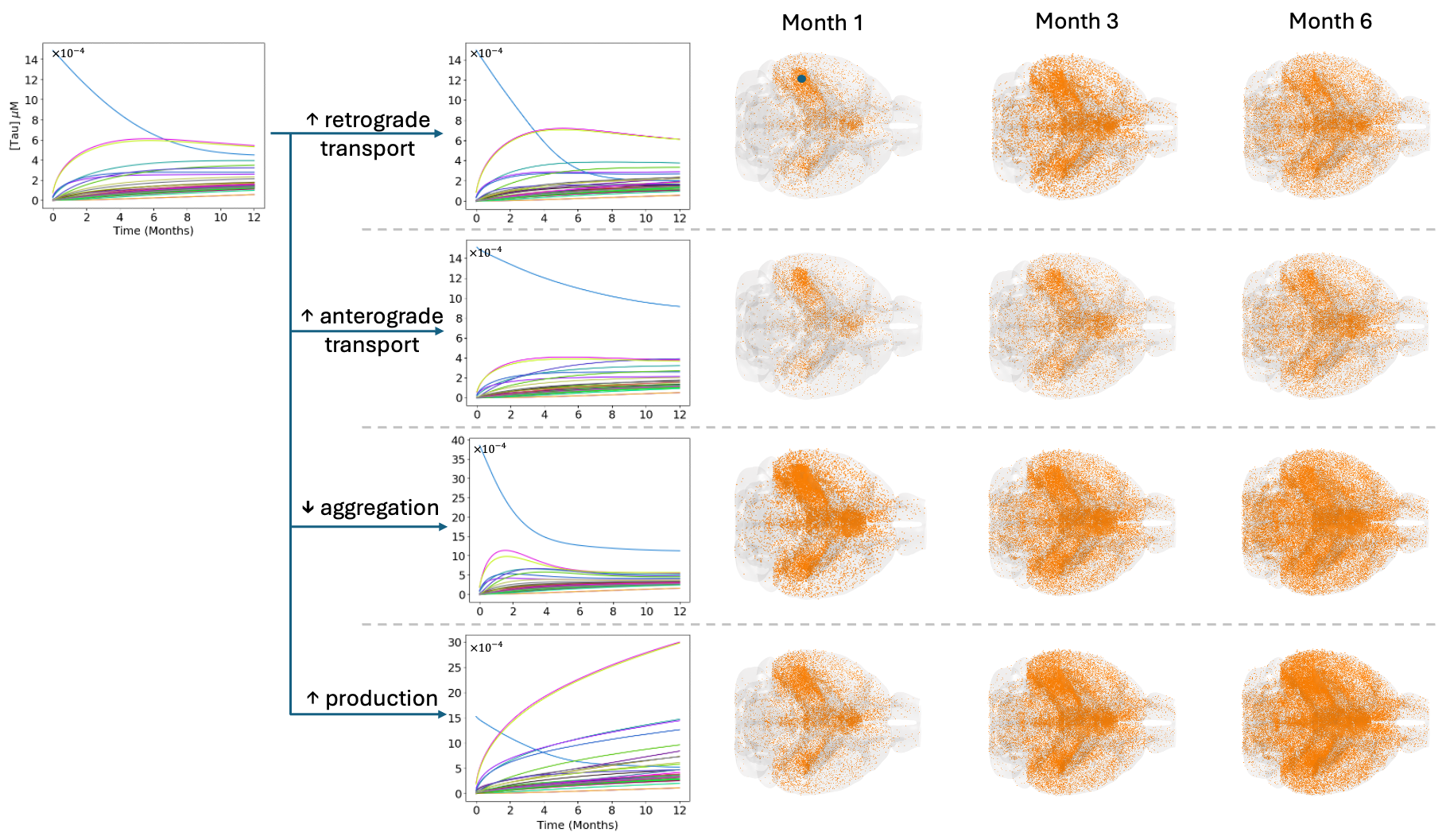}
    \caption{Tau-BNO predictions illustrate the macroscale effects of varying microscopic transport parameters along the mouse structural connectome. Regional tau trajectories (left) are aggregated from the 426-region MCA atlas to the 44-region DS atlas. Glass brain visualizations (right) display spatial tau distributions at selected time points, with increasing orange intensity indicating higher regional tau concentration. The baseline prediction (top left) uses parameter vector $[\lambda_f, \lambda_\gamma, \lambda_\delta, \lambda_\epsilon, \lambda_\mu] = [5\times10^{-4}, 8\times10^{-3}, 10, 10, 2.2]$. Increased anterograde and retrograde transport conditions are generated by modifying $(\lambda_{\delta}, \lambda_{\epsilon})$ to $(100, 10)$ and $(10, 100)$, respectively. Decreased aggregation is modeled with $\lambda_{\gamma} = 1\times10^{-3}$, and increased production with $\lambda_f = 1\times10^{-2}$. The blue dot in the baseline glass brain denotes the initial seeding location.}
    \label{fig:bio-regimes-fig}
\end{figure}

To demonstrate the practical applications of the Tau-BNO model in the study of tau propagation in the brain, we examined the macroscale effects of modulating microscopic transport and reaction parameters in the NTM. Figure~\ref{fig:bio-regimes-fig} shows that altering individual biophysical parameters produces qualitatively distinct spatiotemporal patterns of tau propagation. 
Introducing a retrograde transport bias (top row) accelerates outward spread from the seed region and increases tau accumulation in distal regions, including bilateral CA3. This effect is further amplified when compared directly with anterograde-biased transport (second row), where retrograde dominance yields faster interhemispheric propagation, as visualized in glass brain representations.
Modulating reaction dynamics produces different global and regional signatures. Decreasing the aggregation rate (third row) and increasing tau production (bottom row) both elevate overall tau burden across the brain. However, their spatial effects differ. Increased production rapidly redistributes concentration away from the seed region, which ceases to dominate within the first two months as other regions exhibit near-logarithmic growth. In contrast, reduced aggregation preserves the seed region as the dominant locus of accumulation, while other regions display transient increases followed by plateauing behavior. Modulating tau aggregation kinetics has the additional effect of modulating tau staging as well. The region ordering by tau concentration and regional tau concentration peaks differ by aggregation condition.
Under low aggregation condition simulated here, non-seed brain regions reach peak tau concentration at region-specific time points between 1-3 months post seeding, while the corresponding high-aggregation condition displays a more monotonic increase in regional tau trajectories through 12 months. 

\section{Discussion}
\label{sec:discussion}

Biophysical modeling of tau protein propagation in Alzheimer's disease has traditionally relied on network diffusion approaches that omit microscopic cellular dynamics or on computationally intensive PDE-based frameworks requiring repeated numerical solutions for arbitrary parameter configurations. The recent network transport model (NTM) represents a major leap forward because, unlike earlier passive diffusion models, it incorporates the active axonal transport of tau via molecular motors, the kinetics of protein aggregation and fragmentation, and exosome-mediated trans-neuronal release and uptake. However, modeling these phenomena requires solving large massively coupled PDEs, which has remained outside the realm of practical utility. Recent neural operators show promise for accelerating such simulations, but existing architectures do not incorporate the mathematical structure and biological constraints specific to neural transport systems.

Tau-BNO addresses this limitation through two key design principles. 
First, decoupling the encoding of initial conditions from kinetic parameters allows the model to respect their distinct biological roles: initial conditions specify spatial pathology distributions, whereas kinetic parameters govern transport and diffusion processes operating on different scales. 
Second, our Directed Graph Operator accounts for the directional asymmetry inherent in neural pathways. By fusing multiple symmetric graph projections, it approximates directed transport while maintaining computational efficiency, 
thereby capturing spatiotemporal propagation patterns more faithfully.

\subsection{Tau-BNO provides a rapid and accurate proxy model of tau spread}

The design choices behind Tau-BNO yield substantial performance gains. 
Tau-BNO achieves a 32\% reduction in RMSE compared with state-of-the-art neural operators, with ablation studies revealing that the decoupled architecture with the Directed Graph Operator provides a further 12\% improvement over these baselines.
Critically, the model maintains high accuracy under extreme parameter regimes, suggesting that it has learned a stable approximation of the underlying solution operator rather than overfitting specific training configurations. These findings underscore a broader principle for operator learning in biological systems: respecting the mathematical and structural constraints of the underlying dynamics can be as important as architectural design choices. 

Tau-BNO provides a means to more readily model tau spread in the brain as defined by NTM dynamics, removing the need for extensive computational resources required when using numeric approaches to solve the associated system of PDEs. Once trained, longitudinal tau simulations can be generated within seconds on standard hardware, compared to hours of computation using conventional PDE-based solvers.
This acceleration makes large-scale parameter exploration and sensitivity analysis computationally feasible.
Beyond efficiency, Tau-BNO enables systematic investigation of how microscopic biophysical parameters shape macroscale tau propagation. 
By supporting rapid forward simulations across diverse parameter regimes, the framework facilitates analysis of transport asymmetries, aggregation dynamics, and uptake processes in relation to emergent spatial pathology patterns. 
Although developed within the NTM framework, the broader Brain Neural Operator paradigm is extensible to other connectome-based models of neurodegeneration.

\subsection{Tau-BNO produces novel mechanistic insights}

Backed by the computational efficiency provided by Tau-BNO, analysis of aggregation dynamics reveal that the tau aggregation rate plays a critical role in the global spread rate, staging, and overall burden of soluble tau in the brain, as shown in Figure \ref{fig:bio-regimes-fig}. 
A decreased aggregation rate results in a more rapid global spread of tau from the seed region to the rest of the brain, which can be biologically explained by immobile aggregated tau polymers sequestering the amount of tau monomers available to spread, thus slowing the overall rate of tau spread.
The effect is not as straightforward as simply modulating the time scale of tau spread, as tau staging is shown to vary by aggregation conditions as well. 
%Under low aggregation condition simulated here, non-seed brain regions reach peak tau concentration at region-specific time points between 1-3 months post seeding, while the corresponding high-aggregation condition displays a more monotonic increase in regional tau trajectories through 12 months. - MOVE TO RESULTS
%In addition, the region ordering by tau concentration differs by aggregation condition. - MOVE TO RESULTS 
The overall soluble tau burden is shown to increase under the low aggregation rate condition, which is expected since a lower aggregation rate favors a larger proportion of soluble tau monomers in the aggregation-fragmentation equilibrium between tau monomers and polymers.

Tau staging and global spread also demonstrate a critical dependence on directional tau transport dynamics. See Figure \ref{fig:bio-regimes-fig}. Retrograde-biased transport (against axonal polarity) regimes demonstrate a faster global and inter-hemispheric tau spread in the brain compared to anterograde-biased transport regimes (in the direction of axonal polarity). The region ordering by tau concentration throughout the simulations also changes depending on the direction of tau transport bias. In contrast, increased tau production rate has its greatest influence on the global spread rate and burden of tau, but not region ordering (with the exception of the seed region). 
%Increasing monomeric tau production predictably increases the overall tau burden over time, which in turn increases the global spread rate. - REMOVE/MOVE TO RESULTS
Together, these results highlight the importance of microscopic tau dynamics in capturing spatiotemporal patterns of tau progression that contemporary diffusion based models fail to capture.
In addition, these analyses highlight Tau-BNO’s capacity to reproduce mechanistically distinct propagation signatures arising from specific microscopic transport and reaction parameters, enabling previously intractable systematic exploration of how cellular-scale processes shape large-scale tau dynamics.

\subsection{Utility, Impact and Generalizations}

The accurate performance of Tau-BNO makes it an attractive surrogate model. Its practical utility lies in finally rendering large-scale parameter inference feasible. This will allow researchers to rapidly fit similar models to empirical data to extract patient-specific system coefficients, evaluate heterogeneous disease trajectories, and explore how specific cellular mechanisms could be targeted by future therapeutics. By successfully approximating complex PDE systems without losing biological interpretability, this work showcases the transformative value of  deep learning to serve as surrogate for intractable biophysical models. Most importantly, the surrogate model is capable of producing new insights and generating new hypotheses. Ultimately, the Tau-BNO framework provides a scalable computational foundation for investigating broader neuroscientific processes, such as axonal transport dynamics and network-mediated disease spread, paving the way for rapid, personalized therapeutic targeting in neurodegenerative diseases. The proposed Tau-BNO is intended to be the centerpiece of an emerging effort to create \emph{in silico} models of neurodegenerative processes, that will serve as a computational workbench for future investigators. 

\subsection{Related Work}

Neural operator architectures provide a principled framework for learning mappings between infinite-dimensional function spaces. DeepONet \citep{lu2021learning} adopts a branch–trunk architecture that encodes input functions and spatiotemporal coordinates separately, learning the solution operator in accordance with universal operator approximation theory \citep{chen1995universal}. The Fourier Neural Operator (FNO) \citep{li2020fourier} performs global spectral convolutions using fast Fourier transforms \citep{cooley1965algorithm}, enabling efficient modeling of long-range interactions. Wavelet-based variants, including the Multiwavelet-based operator model (MWT) \citep{gupta2021multiwavelet} and the Wavelet Neural Operator (WNO) \citep{tripura2023wavelet}, introduce multi-resolution representations that jointly capture spatial and frequency structure \citep{farge1992wavelet}. The Localized Neural Operator (LNO) \citep{liu2024neural} augments operator learning with differential kernels to strengthen local spatial modeling.
While these approaches have demonstrated strong performance for PDE systems defined on structured grids, extending operator learning to irregular geometries and network-like domains remains challenging. Graph-based neural operators address this limitation by incorporating message-passing mechanisms from graph neural networks \citep{bruna2013spectral,wu2020comprehensive}. In particular, the Graph Neural Operator (GNO) \citep{li2020neural} and the Geometry-Informed Neural Operator (GINO) \citep{li2023geometry} replace kernel integration with graph-based propagation, further leveraging Nyström approximations \citep{nystrom1930praktische} to support operator learning on irregular meshes and complex topologies. These developments have broadened the applicability of neural operators to systems defined on non-Euclidean domains.

Within computational neuroscience, graph neural networks (GNNs) have been widely applied to brain connectomics. For example, GNN-based models have been used to study the structure–function relationship between structural connectivity (SC) and functional recordings derived from functional magnetic resonance imaging (fMRI) \cite{tang2025graph, xia2025interpretable, wein2021graph, mohammadi2024graph} and electroencephalography (EEG) \cite{mohammadi2024graph}. Such approaches often achieve competitive predictive performance relative to neural mass models \cite{david2003neural,moran2007neural, byrne2020next}, though they typically operate as statistical models without explicitly encoding mechanistic transport processes. Similarly, GNNs have been employed to forecast tau propagation on structural connectomes \cite{balaji2022graph, dan2024tauflownet}, yet these models do not incorporate the biophysical reaction–transport dynamics underlying mechanistic frameworks such as the NTM.
Neural operators have more recently been explored in neuroscience applications, including Neural Koopman Operators for modeling brain dynamics and cognitive states \cite{zhou2025understanding, gallos2024data, marrouch2020data}, as well as operator-learning approaches for biomechanical modeling of traumatic brain injury \cite{agarwal2025real}. However, operator-based frameworks that explicitly integrate connectome topology with mechanistic transport dynamics remain underdeveloped in neurodegenerative disease modeling. Tau-BNO contributes to this emerging direction by combining operator learning with biologically grounded graph structure tailored to neural transport systems.

\subsection{Limitations and Future Directions}
% Limitations and future directions
Several limitations warrant careful consideration. 
First, our current implementation is trained on mouse connectome data, and the Directed Graph Operator encodes topological priors specific to murine neural architecture. 
The substantial anatomical and network-level differences between mouse and human brains pose challenges for direct translation to human applications. 
In addition, mouse datasets generally specify ground truth tau seed regions, which are unknown in tau-PET derived human datasets.
Future work will therefore prioritize adapting Tau-BNO using human connectome datasets and clinically derived measurements. 
Second, our current framework addresses the forward problem of predicting dynamics from system parameters, whereas clinical applications often require solving the inverse problem—inferring kinetic parameters from observed spatiotemporal tau distributions. 
Developing an inverse modeling framework would transform Tau-BNO from a simulation tool into a diagnostic instrument capable of extracting patient-specific system parameters from monitoring data. 
This could enable earlier detection of pathological changes and more precise characterization of disease subtypes.
Third, our validation relies on synthetic data generated from mechanistic models rather than direct comparison with clinical observations. Although these models incorporate established biophysical principles, discrepancies between simulated and real-world disease progression may limit translational applicability. Integrating Tau-BNO with longitudinal clinical datasets therefore represents a critical next step. The computational speed and accuracy with which Tau-BNO approximates the NTM make large-scale empirical inference feasible, enabling parameter estimation and model fitting that would otherwise be computationally prohibitive. Applying the NTM to observational data using Tau-BNO as a surrogate is an active area of ongoing work.
\section{Biophysical Tau Transport Model Theory}\label{sec:NTM}

The Network Transport Model is a biophysical model defined as a set of partial differential equations outlined in \cite{barron2026biophysically} that explain the macroscopic region-level progression of tau in the brain as a process arising from microscopic cellular-level tau dynamics that include tau aggregation and fragmentation, diffusion and transport, cellular uptake and release, and spontaneous production processes. The NTM thus provides a link between the observed region-level patterns of tau spread and the biophysical cellular mechanisms governing tau behavior. 

\subsection{Single Neuron Dynamics}
 
Single cell tau dynamics are described by a compartmental model of a neuron. This compartmental model draws from single cell tau dynamics introduced by Torok et al. \cite{torok2021emergence}. 
%In this model, a single neuron is separated into three compartments along a one dimensional axis defined on $x \in (0,L)$ as follows:
%\begin{equation}
 %   \begin{cases}
  %      (i)\: \textbf{Presyn. SD}\text{: presynaptic somatodendritic compartment},\: x \in (0,x_1),\\
%        (ii)\: \textbf{AIS}\text{: axon initial segment},\: x \in (x_1,x_2), \\
 %       (iii)\: \textbf{Axon}\text{: axonal compartment},\: x \in (x_2,L)
 %   \end{cases}\label{eq:compartments}
%\end{equation}
%where $L$ is taken to be the length of the neuron, and $0 < x_1 < x_2 < L$. 
Within each compartment of the single neuron model are three primary biophysical processes that govern tau pathology: (1) diffusive spread and active transport; (2) aggregation and fragmentation of soluble tau monomers and aggregated insoluble tau polymers; and (3) the source production of soluble tau in seed regions of the brain. These dynamics are described by the following system of partial differential equations (PDEs):
\begin{equation}
\begin{cases}
\varphi n_t = 
\big(\underbrace{a(x)\,n_x}_{\text{diffusion}} + 
\underbrace{h(x,m,n)}_{\text{active transport}} \big)_{x} + 
\Gamma(m,n) + 
\underbrace{F_{edge}}_{\text{production}},\\[6pt]
\varphi m_t = -\Gamma(m,n), \\[6pt]
\underbrace{\Gamma(m,n) = \beta m  - \gamma_1 n^2 -\gamma_2mn}_{\text{aggregation \& fragmentation}}
\end{cases}
\label{eq:microscale_torok-intro}
\end{equation}
where \(n(x,t)\) and \(m(x,t)\) are soluble and insoluble tau concentrations, respectively, $\varphi$ is a scale factor that describes the ratio between slow and fast time scales different processes act on, $\Gamma(m,n)$ is the conversion rate between soluble and insoluble tau due to aggregation and fragmentation modulated by the $\gamma_1$ and $\gamma_2$ aggregation rate parameters and the $\beta$ fragmentation rate parameter, $a(x)$ is the compartment-specific diffusivity modulating the rate of tau diffusion, and $F_{edge}$ denotes a source production term for soluble tau.
%\begin{equation}
%a(x)=
%\begin{cases}
%D & x \in \text{Presyn. SD} \\
%\lambda_1 D & x \in \text{AIS} \\
%fD & x \in \text{Axon}
%\end{cases}
%\end{equation}
Active transport dynamics are denoted by $h$ and defined as follows:
\begin{equation}
%\begin{cases}
 h(x, m, n)=\begin{cases}
-(1-f)(v_a(1+\delta n)(1-\varepsilon m)-v_r))n &  x \in \text{Axonal Compartment} \\
 0 & \text{  otherwise}\;.
\end{cases}\label{eq:active-transport}
\end{equation}
The \(\delta\) and \(\epsilon\) parameters modulate the interaction between tau and the axonal molecular motors, and $v_a$ and $v_r$ are baseline anterograde and retrograde velocities of tau transport.

\subsection{Network Transport Model}

The Network Transport Model (NTM) extends cellular tau dynamics to the whole brain by applying the single neuron compartmental model of tau dynamics outlined by equations \ref{eq:microscale_torok-intro}-\ref{eq:active-transport} to edges of the brain's structural connectome (SC) graph. The SC graph is defined as a weighted directed graph where nodes represent discrete brain regions as defined by a specified atlas parcellation, and edges represent long range white matter tract neurons interconnecting brain regions weighted white matter tract connectivity strength that is empirically determined from diffusion magnetic resonance imaging (dMRI) tractography in humans and viral tracing in mice.
%The SC graph is defined as follows:
%\begin{equation}\label{weights-intro}
%c(P_i,P_j)=
%\begin{cases}
%c_{ij}>0 & \text{if $P_i$ and $P_j$ are connected by white matter tracts}\;,  \\
%0 & \text{otherwise}\;.
%\end{cases}
%\end{equation}
%where the connectivity edge weights $c_{ij}$ are the ``strength" of the white matter connections between regions $P_i$ and $P_j$ that are empirically determined from diffusion magnetic resonance imaging (dMRI) tractography in humans and viral tracing paired with dMRI in mouse. 
To model tau dynamics on the brain's SC graph, each graph node $P_i$ represents the extracellular space in region $i$, and each directed edge $e_{ij}$ represents the intracellular space of the neurons connecting nodes $P_i$ to $P_j$. 

The exchange of tau between intracellular and extracellular spaces as modulated by uptake and release mechanisms is modeled as a process that occurs on the boundary between SC graph nodes (representing extracellular space in region $P_i$) and edges (representing intracellular space of white matter tract neurons connecting brain regions). For SC graph edge $e_{ij}$, the tau fluxes $J_{ij}$ at nodes $P_i$ and $P_j$ are given by:  
\begin{equation}
\begin{cases}
J_{ij}(P_i)=-\mu_{i,1} n_{ij}(P_i) +\mu_{i,2}N_i,\\
J_{ij}(P_j)=\mu_{j,1} n_{ij}(P_j) -\mu_{j,2}N_j.
\end{cases}
\end{equation}
where $\mu_{i,1}$ and $\mu_{i,2}$ are release and uptake rate parameters between edge $e_{ij}$ and node $P_i$, $\mu_{j,1}$ and $\mu_{j,2}$ are release and uptake rate parameters between edge $e_{ij}$ and node $P_j$, $N_i$ and $N_j$ are the extracellular concentrations of soluble tau at nodes $P_i$ and $P_j$, and $n_{ij}(P_i)$ and $n_{ij}(P_j)$ are the concentrations of intracellular soluble tau at boundary points along $e_{ij}$ with nodes $P_i$ and $P_j$. 

The NTM then captures the concentrations of extracellular soluble and insoluble tau at node $P_i$, denoted $N_i$ and $M_i$ respectively, as follows:
\begin{equation}
\label{eqn:sys nodes} 
\begin{cases}
\phi M_i'=-\Gamma(M_i,N_i) \\
\phi N'_i=\frac 1{V_i}\underbrace{\sum_{j\ne i} 
\left(-c_{ij}J_{ij}(P_i) +c_{ji}J_{ji}(P_i)\right)}_{\text{incoming tau flux into node } P_i}+\Gamma(M_i,N_i)
\end{cases}
\end{equation}
where $V_i$ denotes the volume of region $P_i$, $c_{ij}$ is the connectivity edge weight of SC graph edge $e_{ij}$, and $\phi << 1$ is a small scale constant between slow and fast time scales on which different modeled processes occur.
%Tau production is captured in the edge dynamics, where $F_{edge,ij}$ is the edge $e_{ij}$ specific production rate seen in equation \ref{eq:microscale_torok-intro} defined as follows:
%\begin{equation}\label{F}
%F_{edge,ij}=
%\begin{cases}0  &\text{if }i,j \notin I_{seed}\\
%const &\text{if }i \lor j \in I_{seed},\\
%\end{cases}
%\end{equation}
%where $I_{seed}$ is a set of nodes corresponding to brain regions where tau pathology is assumed to originate from.

\subsection{Quasi Static Approximation}

The quasi-static approximation of the NTM leverages the fact that different biophysical processes mediating tau progression occur on vastly different time scales, allowing for fast-acting processes to reach a steady state before affecting slower acting tau dynamics in an appreciable way. This steady state approximation allows the NTM to be reformulated as a series of ordinary differential equations (ODEs) to solve the governing system of PDEs, which makes generating simulations of the NTM possible, but still extremely computationally intensive and time consuming. The equations governing tau dynamics of the quasi-static NTM on SC nodes are given by:
\begin{equation}
\begin{cases}
M_i=\frac {\gamma_1 N_i^2}{\beta-\gamma_2 N_i} \\
\left(1+\frac{2\beta-\gamma_2 N_i}{(\beta-\gamma_2 N_i)^2}\gamma_1 N_i \right)N_i'=\frac 1{V_i}\sum_{j}\big(c_{ij}\left(\hat\mu_{i,1} n_{ij}(0,t) - \hat \mu_{i,2}N_i(t)\right) 
+c_{ji}
\left(\hat \mu_{i,1} n_{ji}(L_{ji},t) - \hat \mu_{i,2}N_i(t)\right)
\big)\\
\end{cases}\label{eq:quasi-static-node}
\end{equation}
%where the following approximation is made for edgewise dynamics:
% \begin{equation}
% \begin{cases}
% \Gamma(m_{ij},n_{ij})=0 \, \Leftrightarrow\, m_{ij}=g(n_{ij})=\frac {\gamma_1 n_{ij}^2}{\beta-\gamma_2 n_{ij}}&\text{in }(0,L)\\
% %J_{ij}=-a(x) (n_{ij})_x+h(x,n_{ij})\\
% \partial_xJ_{ij}=(F_{edge})_{ij} &\text{in }(0,L).\\
% \end{cases}\label{eq:qs-approx}
% \end{equation}
where $\hat{\mu}_k=\frac{\mu_k}{\phi}$. For a detailed explanation of the quasi-static NTM and numeric methods used to generate simulations, the reader is directed to \cite{barron2026biophysically}.

\section{Methods} \label{sec:theory}
% TODO: Add model architecture Figure.
In this section, we first formalize the problem and then present the proposed Brain Neural Operator. An overview of the method is 
shown in Figure ~\ref{fig:Tau-BNO}.

\subsection{Operator Learning Framework}

\subsubsection{Problem Formulation}
\label{sec:problem-definition}

We consider the spatiotemporal evolution of tau transport over a spatial domain $\Omega \subset \mathbb{R}$ and time interval $\mathcal{T}=[0, T]$. 
Let ${u}: \Omega \times \mathcal{T} \to \mathbb{R}^{V}$ denote the state field representing the concentrations of $V$ distinct tau species (e.g., soluble and insoluble forms).
The dynamics are governed by a parametric system of PDEs:
\begin{equation}
\label{eq:pde-ib}
\begin{cases}
\partial_t {u}(x,t) = {\Psi}_\tau\big[{u}(x,t), \nabla {u}(x,t); \boldsymbol{\lambda}\big], & (x,t) \in \Omega \times \mathcal{T}, \\
{u}(x,0) = {u}_0(x), & x \in \Omega, \\
\mathcal{B}[{u}](x,t) = 0, & (x,t) \in \partial\Omega \times \mathcal{T},
\end{cases}
\end{equation}
where ${\Psi}$ is a non-linear differential operator parametrized by a vector $\boldsymbol{\lambda} \in \mathbb{R}^p$ (encoding biophysical rates such as production, aggregation, and transport velocities), and ${u}_0$ represents the initial spatial distribution.
%%% simplified Operator Learning
To efficiently predict the system dynamics, we define the exact solution operator $\mathcal{S}: \mathcal{X} \times \Lambda \to \mathcal{U}$ as the mapping $\mathcal{S}({u}_0, \boldsymbol{\lambda}) = {u}(\cdot,\cdot)$, which takes an initial condition and a parameter vector to the corresponding spatiotemporal trajectory.
Our objective is to learn a parametric neural operator $\mathcal{N}_{\theta} \approx \mathcal{S}$ that directly predicts the solution ${u}$ across the biophysical parameter space, thereby bypassing the computational cost of iterative numerical solvers.

% \textbf{Solution operator.}
% Define the solution operator $\mathcal{S}:\;\mathcal{X}\times\Lambda \longrightarrow \mathcal{U},\,\mathcal{S}(u_0,\boldsymbol{\lambda}) = u(\cdot,\cdot),$
% which maps any initial–coefficient pair to the spatiotemporal trajectory on $\Omega\times\mathcal{T}$.
% However, the spaces of initial fields and coefficients can be large; repeatedly solving Eq.~\eqref{eq:pde-ib} for each $(u_0,\boldsymbol{\lambda})$ with classical numerical solvers is computationally intensive. We therefore learn a parametric neural operator \(\mathcal{N}_{\theta}:\mathcal{X}\times\Lambda\to\mathcal{U}\) to approximate the solution operator \(\mathcal{S}\) (i.e., \(\mathcal{N}_{\theta}\approx\mathcal{S}\)).

% >>>>>>>>>>>>>>>>>>>> General Neural Operator <<<<<<<<<<<<<<<<<<<<<

%% neural operator: concise version
\subsubsection{Neural Operator}

We realize \( \mathcal{N}_{\theta} \) as a composition of \emph{multi-kernel} layers that mix pointwise channel transforms with integral and differential kernel operators over the spatial domain. 
Let \(Z^{(\ell)}: \Omega \to \mathbb{R}^{D_\ell}\) be the layer-\(\ell\) feature field. 
Each layer acts as
\[
Z^{(\ell+1)}(x) \;=\; \sigma\!\Big( Z^{(\ell)}(x)\, W_\ell \;+\; \big(\mathcal{K}_\ell[Z^{(\ell)}]\big)(x) \;+\; b_\ell \Big),
\]
where \(W_\ell\in\mathbb{R}^{D_\ell\times D_{\ell+1}}\) is a pointwise linear transform, \(b_\ell\in\mathbb{R}^{D_{\ell+1}}\) is a bias, \(\sigma\) is an activation, and \(\mathcal{K}_\ell\) is a kernel operator acting over \( \Omega \). 

We use an input lifting network \( \mathcal{H}: \mathbb{R}^{D_{\mathrm{in}}}\!\to\!\mathbb{R}^{D_0} \) and an output projection \( \mathcal{P}: \mathbb{R}^{D_{L}}\!\to\!\mathbb{R}^{D_{\mathrm{out}}} \), and set
\[
Z^{(0)} \;=\; \mathcal{H}\big(u_0,\boldsymbol{\lambda}\big), 
\qquad
\mathcal{N}_{\theta}(u_0,\boldsymbol{\lambda}) \;=\; \mathcal{P}\big(Z^{(L)}\big).
\]

Collecting learnable parameters from neural operator $\mathcal{N}_\theta$, we write 
\(
\theta=\big\{\mathcal{H},\mathcal{P},\{W_\ell,b_\ell,\mathcal{K}_\ell\}_{\ell=0}^{L-1}\big\}.
\)

The Tau-BNO framework is composed of specialized operator blocks, each utilizing a distinct kernel configuration to capture multi-scale biological dependencies.
For the FO and QO, which are designed to model continuous spatiotemporal dynamics, we employ a dual-kernel strategy that combines a Fourier integral kernel ($\mathcal{K}^{\mathcal{F}}$) to capture global interaction via spectral convolution with a Differential kernel ($\mathcal{K}^{\mathcal{D}}$) enabling explicit modeling of local reaction dynamics and boundary behavior~\citep{li2020fourier,liu2024neural}.
Conversely, the DGO is tailored to the discrete anatomical connectome, utilizing a Graph integral kernel ($\mathcal{K}^{\mathcal{E}}$) to aggregate features from topological neighbors and model inter-regional diffusion~\citep{li2020neural}.
Formal definitions and mathematical derivations for all kernel operators are detailed in Supplementary Materials~\nameref{sec:KernelDetails}.

\subsubsection{Operator Learning}

To implement the operator learning framework numerically, we discretize the continuous solution $u(x,t)$ onto a finite grid defined by $V$ brain regions and $T$ temporal snapshots.
Let $\mathbf{u} \in \mathbb{R}^{V \times T}$ denote the discrete ground-truth trajectory matrix.
Similarly, let $\widehat{\mathbf{u}} = \mathcal{N}_{\theta}(\mathbf{u}_0, \boldsymbol{\lambda}) \in \mathbb{R}^{V \times T}$ represent the prediction of the neural operator, parametrized by weights $\theta$.
Given a training dataset of $N$ samples $\{(\mathbf{u}_0^{(i)}, \boldsymbol{\lambda}^{(i)}, \mathbf{u}^{(i)})\}_{i=1}^N$, we optimize $\theta$ by minimizing the relative $L^2$ projection error:
\begin{equation}
\label{eq:loss}
\min_{\theta} \mathcal{L}(\theta)
\;=\; \sum_{i=1}^{N}
\frac{\big\|\, \mathbf{u}^{(i)} - \widehat{\mathbf{u}}^{(i)} \,\big\|_{2}}
{\big\|\,\mathbf{u}^{(i)}\,\big\|_{2}}\,,
\end{equation}
where $\|\cdot\|_{2}$ denotes the entry-wise $L^2$ norm, and $\mathcal{N}_{\theta}(\cdot)$ returns a solution prediction in $\mathbb{R}^{V\times T}$. 
This objective promotes consistent approximation of the underlying solution operator $\mathcal{S}$ across varying initial conditions and biophysical parameter regimes.

\subsection{Tau-BNO Architecture}

The Tau-Brain Neural Operator (Tau-BNO) is designed to simulate tau transport dynamics by explicitly disentangling local kinetics from global transport. 
As illustrated in Figure~\ref{fig:Tau-BNO}, the framework decomposes the solution operator into three specialized modules.
Unlike standard approaches that conflate these factors into a unified latent representation, Tau-BNO processes kinetics and initial condition information via independent pathways before fusing them. Formally, the composite operator $\mathcal{N}_\theta$ is defined as:
\begin{equation}
    \mathcal{N}_\theta 
    = \mathcal{P} \circ \mathcal{G} \circ (\mathcal{R} \odot \mathcal{Q}) 
    = \mathcal{P} \circ \mathcal{G} \circ (
    (\widetilde{\mathcal{R}}  \circ \mathcal{H}_\mathcal{R}) \odot (\widetilde{\mathcal{Q}} \circ \mathcal{H}_\mathcal{Q})),
\end{equation}
where \( \mathcal{N}_\theta \) is our Tau-BNO model, \( \mathcal{H} \) and \( \mathcal{P} \) are the lifting neural network and solution trajectory projector. The operator \( \mathcal{G} \) represents the DGO, \( \mathcal{R} \) is the FO, and \( \mathcal{Q} \) is the QO.

% \paragraph{Contributions.}
% \begin{enumerate}
%     \item We utilize two neural operators (FO and QO) to encode regional temporal dynamics and a directed graph operator to propagate cross-region transport, thereby injecting targeted inductive biases that support reliable spatiotemporal prediction.
%     \item We factor the architecture into a Function Operator for regional kinetics and a Query Operator for instantaneous regional conditions, yielding region-specific temporal dynamics.
%     \item We leverage an experimentally derived directed brain network and a directed graph operator; to promote smooth outputs while preserving directional asymmetry in graph convolution, We fuse three undirected surrogates (first-order symmetrized, two-order in neighborhood, and two-order out neighborhood) to approximate directed interactions and produce direction-respecting spatiotemporal trajectory predictions.
% \end{enumerate}

\subsubsection{Function Operator}
To extract regional kinetics and better characterize the local dynamical behavior within each brain region, 
we introduce a \emph{Function Operator} (FO) that learns a mapping from the initial condition \(\mathbf{u}_0\in\mathcal{X}(\Omega;\mathbb{R})\) and system parameters \(\boldsymbol{\lambda}\in\Lambda\) to regional kinetic representations. For a fixed \(\boldsymbol{\lambda}\), the parameters are broadcast to all regions, promoting the sharing of system parameters while allowing regional specific kinetic adaptation.
Formally, the operator is defined as
\[
\mathcal{R}: \mathcal{X}\times \Lambda(\Omega;\mathbb{R}^{p+1}) 
\;\longrightarrow\; 
\mathcal{Y}_{\mathcal{R}}(\Omega;\mathbb{R}^{D}),
\qquad
\boldsymbol{r}(x):=(\mathcal{R}[\mathbf{u}_0,\boldsymbol{\lambda}])(x)\in\mathbb{R}^{D}.
\]
% where $\Omega$ is the discrete set of n brain regions, and interpret $x\in\Omega$ as a region index, \(p\) denotes the number of tau-transport system parameters, and \(D\) represents the hidden feature dimension\footnote{Throughout, we fix the hidden dimension to $D$ for all layers.}.
where $\Omega$ denotes the set of $n$ brain regions indexed by $x$, $p$ represents the number of tau-transport parameters, and $D$ is the hidden dimension (fixed across all layers).
The Function Operator is constructed as a composition of an input lifting network, a sequence of function kernel layers, and an output projection:
% $$
\begin{equation}\label{eq:function-operator}
  \mathcal{R}
  \;=\;
  \mathcal{P}_{\mathcal{R}}
  \circ
  \mathcal{M}_{\mathcal{R},L-1}
  \circ
  \cdots
  \circ
  \mathcal{M}_{\mathcal{R},0}
  \circ
  \mathcal{H}_{\mathcal{R}},
\end{equation}
% $$
where \(\mathcal{H}_{\mathcal{R}}\) denotes the input lifting module, 
\(\mathcal{P}_{\mathcal{R}}\) is the output projection, 
and each \(\mathcal{M}_{\mathcal{R},\ell}\) represents a function kernel layer, and the operator consists of $L$ such layers.

Each layer \(\mathcal{M}_{\mathcal{R},\ell}\) integrates two complementary kernels:
a \emph{Fourier integral kernel} 
\(\mathcal{K}_{\mathcal{R}}^{\mathcal{F}}\) 
inspired by the FNO~\citep{bahouri2011fourier} as in Eq.~\eqref{eq:fourier-kernel} 
and a differential kernel 
\(\mathcal{K}_{\mathcal{R}}^{\mathcal{D}}\) following the LNO~\citep{liu2024neural} formulation in Eq.~\eqref{eq:differential}.
These two kernels are combined additively, resulting in
\begin{equation}\label{eq:function-layer}
  \mathcal{M}_{\mathcal{R},\ell+1}[Z](x)
  \;=\;
  \sigma\!\Big(
    Z^{(\ell)}(x)W_{\mathcal{R},\ell}
    +
    \big(\mathcal{K}_{\mathcal{R},\ell}^{\mathcal{F}}[Z^{(\ell)}]\big)(x)
    +
    \big(\mathcal{K}_{\mathcal{R},\ell}^{\mathcal{D}}[Z^{(\ell)}]\big)(x)
    +
    b_{\mathcal{R},\ell}
  \Big),
\end{equation}
% where \(\sigma(\cdot)\) denotes a nonlinear activation function,
% \(W_{\mathcal{R},\ell}\) is a pointwise linear transformation,
% and \(b_{\mathcal{R},\ell}\) is a bias term.
where \(\sigma\) is the activation function, and \(\{W_{\mathcal{R},\ell}, b_{\mathcal{R},\ell}\}\) represent the learnable affine parameters.

This complementary design ensures that the resulting representation effectively captures global, long-range dependencies while simultaneously resolving local dynamics.

\subsubsection{Query Operator}

To explicitly condition the predicted dynamics on the initial condition, we introduce the \emph{Query Operator} (QO).
Unlike the FO, the QO derives a region-specific feature map solely from the $\mathbf{u}_0$. 
This disentanglement ensures that the learned dynamics are modulated by the precise initial condition, thereby preventing the washout of initial conditions during temporal evolution and improving the identifiability of state-dependent trajectories.
Given an initial condition \(\mathbf{u}_0\in\mathcal{X}(\Omega;\mathbb{R})\), the QO produces a regional condition embedding:
\[
\mathcal{Q}:\ \mathcal{X}(\Omega;\mathbb{R})
\ \longrightarrow\
\mathcal{Y}_{\mathcal{Q}}(\Omega;\mathbb{R}^{D}),
\qquad
\boldsymbol{q}(x):=(\mathcal{Q}[\mathbf{u}_0])(x)\in\mathbb{R}^{D}.
\]
Architecturally, \(\mathcal{Q}\) mirrors the Function Operator but uses a $\mathcal{H_\mathcal{Q}}:\mathbb{R}\rightarrow \mathbb{R}^{D}$ at the input, focusing the capacity on extracting regional cues from \(\mathbf{u}_0\) rather than predicting kinetics.
We realize \(\mathcal{Q}\) as a composition of an input lifting, a stack of \(L\) query layers, and an output projection:
% $$
\begin{equation}\label{eq:query-operator}
  \mathcal{Q}
  \;=\;
  \mathcal{P}_{\mathcal{Q}}
  \circ
  \mathcal{M}_{\mathcal{Q},L-1}
  \circ \cdots \circ
  \mathcal{M}_{\mathcal{Q},0}
  \circ
  \mathcal{H}_{\mathcal{Q}},
\end{equation}
% $$
where \(\mathcal{H}_{\mathcal{Q}}\) is a \emph{linear} lifting at each $x$ 
and \(\mathcal{P}_{\mathcal{Q}}\) represents the output projection that produces the regional condition embedding.
Each layer \(\mathcal{M}_{\mathcal{Q},\ell}\) follows the same architecture as described in Eq.~\eqref{eq:function-layer}.

% \textbf{Hadamard product with the Function Operator.}
Let \(\mathcal{R}[\mathbf{u}_0,\boldsymbol{\lambda}](x)\in\mathbb{R}^{D}\) denote the parameter driven regional kinetics produced by the FO.
We combine the two pathways through a pointwise gated interaction based on the Hadamard (element wise) product:
\begin{equation}\label{eq:query-fusion}
  \boldsymbol{f}(x)
  \;:=\;
      \mathcal{R}[\mathbf{u}_0,\boldsymbol{\lambda}](x)
      \odot
      \mathcal{Q}[\mathbf{u}_0](x),
\end{equation}
yielding the regional temporal dynamics \(\boldsymbol{f}(x)\).
This design lets the initial condition embedding \(\mathcal{Q}[\mathbf{u}_0]\) modulate the kinetics, thereby preserving identifiability while enabling expressive conditioning.

\subsubsection{Directed Graph Operator}
To model interregional tau transport and diffusion along the brain connectome, we employ a modified graph kernel that enables anisotropic, attribute aware message passing on the regional state \(\boldsymbol{f}(x)\). 
This kernel captures direction dependent connectivity while modulating interactions by node and edge covariates, thereby driving the spatiotemporal evolution of regional dynamics.

To obtain a numerically stable yet expressive operator, we instantiate this kernel via \emph{spectral graph convolutions} evaluated on multiple \emph{symmetric surrogate graphs} constructed from the directed connectome. 
This choice is motivated by the classical spectral graph convolution framework \citep{kipf2016semi}, which presumes a \emph{symmetric} graph structure so that the Laplacian operator is self adjoint with an orthonormal eigenbasis.
For a directed adjacency $\mathbf{A}\in\mathbb{R}_{\ge 0}^{n\times n}$, the associated Laplacians are generally non-symmetric, leading to possibly non-orthogonal eigenvectors.
% \textbf{Graph proximity.} 
Accordingly, we consider three symmetric surrogates that capture distinct structural proximities within the directed connectome \citep{tong2020directed}. 
The \emph{first-order symmetry proximity} preserves the direct connectivity of the original graph, while the \emph{second-order in-degree} and \emph{second-order out-degree proximities} encode second-order inbound and outbound connection patterns, respectively. 

Formally, we define the symmetric, in-degree, and out-degree proximity matrices as $\mathbf{A}_{\mathrm{sym}} = (\mathbf{A}+\mathbf{A}^{\top})/2$, $\mathbf{A}_{\mathrm{in}} = \mathbf{A}^{\top}\mathbf{D}_{\mathrm{out}}^{-1}\mathbf{A}$, and $\mathbf{A}_{\mathrm{out}} = \mathbf{A}\mathbf{D}_{\mathrm{in}}^{-1}\mathbf{A}^{\top}$, respectively, where $\mathbf{D}_{\mathrm{in/out}}$ are the corresponding degree matrices.
To ensure numerical stability and spectral smoothness, we apply the standard symmetric normalization~\citep{kipf2016semi} to each surrogate: \begin{equation} \label{eq:gcn-sym-norm}
\widehat{\mathbf{A}}
:= \widetilde{\mathbf{D}}^{-1/2}
   \widetilde{\mathbf{A}}
   \widetilde{\mathbf{D}}^{-1/2},
\qquad
\widetilde{\mathbf{A}} = \mathbf{A} + \mathbf{I},
\qquad
\widetilde{\mathbf{D}} := 
\mathrm{diag(\widetilde{\mathbf{A}}\mathbf{1})}.
% \sum_j \widetilde{\mathbf{A}}_{\mathrm{sym},ij}.
\end{equation}

% \paragraph{Symbolic Formulation of the Directed Graph Operator.}
% \textbf{Spatiotemporal graph propagation layer.} 
Under this formulation, the spatiotemporal dynamics of tau propagation are modeled by a DGO as:
% $$
\begin{equation}\label{eq:directed-graph-operator}
  \mathcal{G}
  \;=\;
  \mathcal{P}_{\mathcal{G}}
  \circ
  \mathcal{M}_{\mathcal{G},L-1}
  \circ \cdots \circ
  \mathcal{M}_{\mathcal{G},0}
  \circ
  \mathcal{H}_{\mathcal{G}},
\end{equation}
% $$
where $\mathcal{H}_{\mathcal{G}}$ encodes the input embedding and spatial transformation, 
$\mathcal{M}_{\mathcal{G},\ell}$ denotes a message passing module employing three distinct kernel functions corresponding to the three proximity undirected graphs ($\mathbf{A}_{\mathrm{sym}}, \mathbf{A}_{\mathrm{in}}, \mathbf{A}_{\mathrm{out}}$), 
and $\mathcal{P}_{\mathcal{G}}$ represents the fusion projector that integrates the multi kernel responses into a unified representation of the tau propagation dynamics.

Formally, the operator induces the mapping
\[
\mathcal{G}:\ 
\mathcal{Y}_{\mathcal{R}}(\Omega;\mathbb{R}^{D})
\odot
\mathcal{Y}_{\mathcal{Q}}(\Omega;\mathbb{R}^{D})
\longrightarrow
\mathcal{Y}_{\mathcal{G}}(\Omega;\mathbb{R}^{D}),
\qquad
\boldsymbol{g}(x) := (\mathcal{G}[\boldsymbol{f}])(x) \in \mathbb{R}^{D},\ \ \text{for } x\in\Omega.
\]

% \paragraph{Directed Graph Operator Layer.}
Let $\{\mathrm{sym},\,\mathrm{in},\,\mathrm{out}\}$ denote the indices of the
symmetric surrogate proximity graphs constructed from the directed adjacency $\mathbf{A}$.
Given node wise features $Z^{(\ell)}:\Omega\to\mathbb{R}^{D}$ at layer $\ell$, 
the message passing module acts on $Z^{(\ell)}$ via three independent branches:
$$
  \big(\mathcal{M}_{\mathcal{G},\ell}[Z]\big)(x)
  \;:=\;
  \left[
  \mathcal{M}_{\mathcal{G}_\mathrm{sym},\ell}[Z](x); \ 
  \mathcal{M}_{\mathcal{G}_\mathrm{in},\ell}[Z](x); \ 
  \mathcal{M}_{\mathcal{G}_\mathrm{out},\ell}[Z](x);
  \right]^{\top}
  \in\; 
  \mathbb{R}^{3 \times D}.
$$

For each $k\in\{\mathrm{sym},\,\mathrm{in},\,\mathrm{out}\}$, 
and $Z_{k}^{(\ell)}(x)$ with initial input $Z_{k}^{(0)}(x) = \mathcal{H}_\mathcal{G}(\boldsymbol{f}(x))$, where $\boldsymbol{f}$ is the regional temporal dynamics from Eq.~\eqref{eq:query-fusion}, 
the $k$-th branch performs a neighborhood aggregation on the normalized
surrogate $\widehat{\mathbf{A}}_{k}$ in Eq.~\eqref{eq:gcn-sym-norm}:
$$
% \begin{equation}\label{eq:branch-node}
  \mathcal{M}_{\mathcal{G}_k,\ell}[Z](x)
  :=\;
  \sigma\!\Big(
      Z_{k}^{(\ell)}(x)\,W^{\mathrm{self}}_{k,\ell}
      \;+\;
      \big(\mathcal{K}^{\mathcal{E}}_{k,\ell}[Z_{k}^{(\ell)}]\big)(x)
      \;+\;
      b_{k,\ell}
  \Big),
% \end{equation}
$$
where $\big(\mathcal{K}^{\mathcal{E}}_{k,\ell}[Z_{k}^{(\ell)}]\big)(x)$ is the graph based integral kernel in Eq.~\eqref{eq:graph-integral-operator}.

% The integral kernel term for each undirected proximity graph is defined nodewise by
% $$
%   \big(\mathcal{K}^{\mathrm{graph}}_{k,\ell}[Z]\big)(x)
%   \;:=\;
%   \int_{U(x)} \kappa_{k,\ell}(x,\xi)\,Z(\xi)\,\mathrm{d}\xi
%   \;\approx\;
%   \sum_{\xi\in U(x)} \widehat{A}_{k}(x,\xi)\, Z^{(\ell)}(\xi)\, W^{\mathrm{msg}}_{k,\ell},
% $$
% where $W^{\mathrm{msg}}_{k,\ell}\in\mathbb{R}^{D\times D}$ transforms the aggregated neighborhood features
% $\sum_{\xi\in U(x)}\widehat{A}_k(x,\xi)\,Z^{(\ell)}(\xi)$.

% \paragraph{Output Fusion Layer.}
% \textbf{Output fusion layer.}
Given the results from three proximity graphs $\mathcal{M}_{\mathcal{G}_\mathrm{sym},L-1}[Z](x)$, 
$\mathcal{M}_{\mathcal{G}_\mathrm{in},L-1}[Z](x)$, and $\mathcal{M}_{\mathcal{G}_\mathrm{out},L-1}[Z](x)$,
we define the output fusion layer as a mapping
$
  \mathcal{P}_{\mathcal{G}}:\ (\Omega;\mathbb{R}^{3 \times D}) \longrightarrow 
  \mathcal{Y}_\mathcal{G}(\Omega;\mathbb{R}^{D}),
$
where the fusion process aggregates the outputs from the three graphs for each node $x\in \Omega$:
% \begin{equation}\label{eq:fusion-process}
$$
  \mathcal{P}_{\mathcal{G}} \big[ \mathcal{M}_{\mathcal{G}_\mathrm{sym},L-1}[Z](x), 
  \mathcal{M}_{\mathcal{G}_\mathrm{in},L-1}[Z](x), \mathcal{M}_{\mathcal{G}_\mathrm{out},L-1}[Z](x) \big]
  \;:=\;
  \frac{1}{3} \sum_{k\in\{\mathrm{sym},\mathrm{in},\mathrm{out}\}} \mathcal{M}_{\mathcal{G}_k,L-1}[Z](x).
$$
% \end{equation}
% final solution
% As a result, the spatiotemporal solution embedding is
% \begin{equation}\label{eq:st-embedding}
%   \boldsymbol{g} \;:=\; \mathcal{G}(\boldsymbol{f}) \ \in\ (\Omega \!\to\! \mathbb{R}^{D}),
%   \qquad
%   \boldsymbol{g}(x) \;=\; \big(\mathcal{G}(\boldsymbol{f})\big)(x),\ \ x\in\Omega .
% \end{equation}
% where $\boldsymbol{f}:\Omega\!\to\!\mathbb{R}^{D}$ denotes the regional temporal dynamics from Eq.~\eqref{eq:query-fusion}.

\subsubsection{Trajectory Projection}
Given the learned spatiotemporal representation, we map it to the target tau trajectory field via a projector
$$
% \begin{equation}\label{eq:projector-map}
  \mathcal{P}:\ \mathcal{Y}_{\mathcal{G}}(\Omega;\mathbb{R}^{D})
  \longrightarrow
  \mathcal{U}(\Omega;\mathbb{R}^{T}),
% \end{equation}
$$
where $T\in\mathbb{Z}$ denotes the length of the target trajectory. 
The operator $\mathcal{P}$ is implemented as a composition of linear layers. 
Consequently, for the representation $\boldsymbol{g}\in\mathcal{Y}_{\mathcal{G}}(\Omega;\mathbb{R}^{D})$ produced by the graph operator, the predicted trajectory is
$$
% \begin{equation}\label{eq:trajectory-output}
  \widehat{\mathbf{u}}
  \;:=\;
  \mathcal{P}[\boldsymbol{g}]
  \;\in\;
  \mathcal{U}(\Omega;\mathbb{R}^{T}),
% \end{equation}
$$
where $\widehat{\mathbf{u}}(x)\in\mathbb{R}^{T}$ gives the predicted time discrete trajectory at location $x$.

% Our training objective utilized the relative loss:
% \begin{equation*}
% \mathcal{L} = \frac{\|\mathbf{u} - \widehat{\mathbf{u}}\|_2^2}{\|\mathbf{u}\|_2^2}
% \end{equation*}
% where $\mathbf{u}$ represents the ground truth trajectory and $\widehat{\mathbf{u}}$ denotes the predicted trajectory. 

\subsection{Tau Transport Trajectory Data Generation}

To train our surrogate for tau propagation dynamics, we generated a comprehensive dataset by numerically solving the governing PDEs across diverse physiological parameter regimes. 
Following the mathematical formulation in \citep{torok2021emergence}, we simulated tau protein concentration dynamics across 426 brain regions over 1,071 distinct combinations of initial conditions and transport parameters, yielding a high dimensional dataset that captures the complex interplay between network topology and biochemical kinetics. All parameters, aside from $\lambda_f$, are randomly sampled from a uniform distribution within a range that is deemed biophysically plausible. Tau production rate $\lambda_f$ is randomly sampled with a 90\% probability from a uniform distribution $U[0,0.001]$ and a 10\% probability from a uniform distribution $U[0.001,0.01]$. This is because the range of $\lambda_f \in [0,0.001]$ is considered the likely range for most empirical datasets of longitudinal tau derived from mouse studies, but a small number of empirical tau mouse datasets show larger tau production over time that can be captured in the range of $\lambda_f \in [0.001, 0.01]$. Larger values of $\lambda_f$ tend to dominate the reaction dynamics, diminishing the relative influence of transport rates. For both of these reasons, this sampling strategy is employed to bias the model training on smaller values of $\lambda_f$.

The SC graph is critical in modeling tau progression because tau has been shown to propagate along white matter tracts in the brain, meaning the SC network topology forms a structural backbone on which tau spread occurs. As such, the NTM models tau spread as a process occurring on the brain's connectivity network. The SC graph represents brain regions, specified by a preselected brain atlas, as graph nodes, and graph edges as neuronal white matter tract connections between regions weighted by the experimentally determined connectivity strength. The connectome used in this study is the mouse mesoscale connectivity atlas (MCA) from the Allen Institute for Brain Sciences \cite{oh2014mesoscale}, obtained using viral tracing methods to determine strength (edge weight) and polarity (edge directionality) of white matter tract connections in the brain. The MCA parcellates each hemisphere of the mouse brain into 213 regions, resulting in an SC graph consisting of 426 regions in the whole mouse brain. However, for visualization purposes, we use a more coarse 44-region parcellation of the mouse brain used as the native space for sampling empirical tau data in Kaufman, et al. \cite{kaufman2016tau} referred to here as the ‘DS’ atlas when plotting tau trajectories (regional concentration over time). Glass brains utilize the finer-grade MCA atlas used in NTM simulations. To convert tau simulations from the 426-region MCA to the 44-region DS atlas for visualization, we compute the volume corrected mean tau concentration of all MCA atlas regions contained in a single DS atlas region at each simulated time point.

The initial tau seeding patterns for the 1,071 NTM simulations used for training were selected to correspond to seed injection regions from 12 experimental mouse studies of tau progression: BoludaCBD \cite{boluda2015differential}, BoludaDSAD \cite{boluda2015differential}, DS4 \cite{kaufman2016tau}, DS6 \cite{kaufman2016tau}, DS6 110 \cite{kaufman2016tau}, DS7 \cite{kaufman2016tau}, DS9 \cite{kaufman2016tau}, DS9 110 \cite{kaufman2016tau}, Hurtado \cite{hurtado2010abeta}, IbaHippInj \cite{iba2013synthetic}, IbaStrInj \cite{iba2013synthetic}, and IbaP301S \cite{iba2015tau}. Seed regions were randomly selected with equal probability from seed regions specified by these 12 studies. The Hurtado study is unique in that mice were not injected with tau, instead the mouse model naturally developed tau, and so no ground truth seed is known. Instead, the seed regions corresponding to this study are selected as the brain regions that contained tau at the earliest measurement (2 months) while maintaining relative tau proportions (measurement intensity) between seed regions. As these studies specify seed regions using different, lower resolution brain atlases than the MCA, the study specified seed region(s) were matched to region(s) in the MCA, generally increasing the number of seed regions under the higher resolution MCA brain atlas. The concentration of tau at each seed region is then determined by aggregation and fragmentation dynamics of the NTM model, described in Equation \ref{eq:microscale_torok-intro} assuming a constant tau seed mass across simulations. This effectively scales seed tau concentrations depending on the sampled value of $\lambda_{\gamma}$. These initial seed tau methods were designed to match known seeding patterns in empirical mouse tau datasets that the NTM model can be applied to study. When applying this model to human tau-PET data with no known tau seeding patterns in future work, the sampling strategy for initial seeded tau will differ to enable our model to generalize across more diverse disease onset patterns observed clinically.

\subsection{Experimental Setup}

All models were implemented in PyTorch 2.8 and trained on NVIDIA RTX A6000 GPUs with 48GB memory. We employed the AdamW optimizer \citep{loshchilov2019decoupled} with model specific learning rates ranging from $2 \times 10^{-4}$ to $1 \times 10^{-3}$ (detailed in Table~\ref{tab:architecture_details}) and a minimum learning rate of $2 \times 10^{-6}$. The learning rate schedule followed cosine annealing \citep{loshchilov2017sgdr} over 1,000 training epochs. 
Complete implementation details are provided in the Supplementary Materials~\nameref{sec:ImplementationDetails}.

\section*{Data and Code Availability}
The data and code supporting the findings of this study will be made publicly available upon acceptance of the manuscript.
% Code for the NTM model is publicly available at https://github.com/Raj-Lab-UCSF/NTM\_ru. 

\section*{Author Contributions}
N.B. and H.R. contributed equally to this work and co-wrote the original draft of the manuscript. 
H.R. designed and implemented the proposed method, conducted the experiments, and contributed to data analysis and visualization. 
N.B. generated the datasets and contributed to data analysis and visualization. 
U.S., D.Y., Y.G., Z.L., and G.Y. contributed to manuscript revision and provided critical feedback. 
M.C. and A.R. supervised the project and provided overall guidance throughout the study.

\section*{Funding}

% N.B. and A.R. have been supported by NIH grants R01AG072753 (A145132), R21AG087921 (A144304), and RF1AG087302 (A144450). 

\section*{Declaration of Competing Interests}
The authors declare no competing interests.

\section*{Acknowledgments}

\section*{Supplementary Material}

\subsection*{Supplementary Methods}

\subsubsection*{Detailed Kernel Formulations}\label{sec:KernelDetails}

\textbf{Integral kernel.}
A integral kernel can be expressed as 
\[
\big(\mathcal{K}_\ell^{\mathrm{int}}[Z^{(\ell)}]\big)(x) \;=\; \int_{\Omega} \kappa_\ell(x,\xi)\, Z^{(\ell)}(\xi)\,\mathrm{d}\xi,
\]
where \(\kappa_\ell:\Omega\times \Omega\to \mathbb{R}^{D_{\ell+1}\times D_\ell}\) is a learnable kernel.
This operator performs global aggregation over the spatial domain, enabling the layer to capture long-range and anisotropic dependencies. 

\textbf{Graph-based integral kernel.}
Following \citep{li2020neural}, we model inter-regional transport and diffusion on the brain network. 
We endow the set of regions $\Omega$ with a graph structure. 
For each node $x\in\Omega$, we specify a finite neighborhood $U(x)\subseteq\Omega$.
Let \(\kappa_{\ell}:\Omega\times\Omega\to\mathbb{R}^{D_{\ell+1}\times D_{\ell}}\) be a learnable kernel that maps source features to target features while encoding connectivity and edge attributes (e.g., orientation and empirically measured tract weights).
Given a feature field $Z^{(\ell)}:\Omega\to\mathbb{R}^{D_\ell}$, the $\ell$-th graph-kernel layer is the integral transform 
\begin{equation}\label{eq:graph-integral-operator}
  \bigl(\mathcal{K}_{\ell}^{\mathcal{E}}[Z^{(\ell)}]\bigr)(x)
  \;:=\;
  \int_{U(x)} \kappa_{\ell}(x,\xi)\,Z^{(\ell)}(\xi)\,\mathrm{d}\xi.
\end{equation}

\textbf{Fourier-based integral kernel.}
In many physical systems governed by partial differential equations (PDEs),
transforming the spatial variables to the Fourier domain via the Fourier transform simplifies both analysis and computation, as differential operators become algebraic in the frequency domain \citep{bahouri2011fourier}.

% Unlike the graph-based kernel, which can precisely characterize the spatial structure of brain networks, 
% the Fourier kernel focuses on capturing temporal coherence and global dependencies through frequency-domain interactions, 
% serving as a complementary component to the subsequent graph-based modeling of spatial relationships.

Unlike the graph-based kernel that focuses on spatial connectivity, the Fourier kernel processes the signal in terms of its time–harmonic components (sine/cosine waves). By assigning learned weights to these components, it enforces temporal smoothness and captures global, long-range patterns, complementing the spatial modeling of the graph-based module.

To efficiently capture translation-invariant and temporally coherent regional dynamics, 
we follow the Fourier Neural Operator \citep{li2020fourier} framework and construct the integral kernel in the spectral domain via the Fourier transform \(\mathcal{F}\):
\begin{equation}\label{eq:fourier-kernel}
  \big(\mathcal{K}_\ell^{\mathcal{F}}[Z]\big)(x)
  \;=\;
  \mathcal{F}^{-1}\!\Big(\,\overline{\mathcal{F}(\kappa_\ell)},\cdot\,\mathcal{F}(Z)\Big)(x),
\end{equation}
where \(\overline{\mathcal{F}(\kappa_\ell)}\) is a learnable spectral multiplier that can be truncated to a predefined finite frequency modes to reduce computational cost and impose spectral regularization.

\textbf{Differential kernel.}
Following the construction in \citep{liu2024neural}, we obtain a differential kernel layer by constraining and scaling a local convolutional kernel so that it converges to a first-order differential operator. 
On an equidistant grid with spacing \(h\) and considering our problem defined on a one-dimensional spatial domain, let the kernel \(K_\ell = (K_\ell)_i\) represent the kernel values at each stride \(i\), with the sum given by \(\overline{K}_\ell := \sum_{i} (K_\ell)_i\). 
We define the first-order differential kernel as:
\[
\big(\mathcal{K}_\ell^{\mathcal{D}}[Z]\big)(x)
\;:=\;
\frac{1}{h}\,\Big(\mathrm{Conv}_{\,K_\ell - \overline{K}_\ell}\,Z\Big)(x).
\]
Then, as \(h \to 0\),
\[
\frac{1}{h}\,\Big(\mathrm{Conv}_{\,K_\ell - \overline{K}_\ell}\,Z\Big)(x)
\;\longrightarrow\;
\nabla Z(x)\,\cdot\, {b}_{\ell},
\]
where \({b}_{\ell}\) defines the learnable kernel moments, with \(\xi\) representing a spatial coordinate in the region near the local coordinate \(x\).

Formally, the differential kernel can be expressed as a convolution as shown in Eq.~\eqref{eq:differential}.
\begin{equation}
    \big(\mathcal{K}_\ell^{\mathcal{D}}[Z]\big)(x)
    \:=\;
    \lim_{h \to 0} \frac{1}{h}\,\Big(\mathrm{Conv}_{\,K_\ell - \overline{K}_\ell}\,Z\Big)(x)
    \:=\;
    \nabla Z(x)\,\cdot\, {b}_{\ell}.
\label{eq:differential}
\end{equation}
The proof is provided in Supplementary~\nameref{sec:diff-proof}.

\subsubsection*{Implementation Details}\label{sec:ImplementationDetails}
% Model Details
Table~\ref{tab:architecture_details} summarizes the architectural configurations for all evaluated models. 
We categorized methods into two groups: \textit{General Architectures} encompassing standard deep learning approaches (MLP, KAN, Neural ODE, Transformer, Mamba, and DeepONet), and \textit{Neural Operators} including specialized PDE-solving architectures (FNO, MWT, WNO, GINO, LNO, and our proposed TauBNO).
To ensure statistical robustness, all experiments were repeated three times with different random seeds, and we report mean performance metrics with standard deviations across runs.

\begin{table}[ht]
\centering
\caption{Architectural specifications and hyperparameters for all evaluated models. L-layer denotes the number of layers; FNO modes indicates the number of Fourier modes retained in frequency domain methods.}
\label{tab:architecture_details}
\small
\begin{tabular}{ccccccccc}
\toprule
\textbf{Type} & \textbf{Method} & \textbf{L-Layers} & \textbf{Hidden Dim} & \textbf{FNO Modes} & \textbf{LR} & \textbf{Dropout} & \textbf{Act.} \\
\midrule
\multirow{6}{*}{\shortstack[c]{General\\Architecture}} 
& MLP & 3 & [32, 128, 256] & --- & $2\times 10^{-4}$ & 0.1 & GELU \\
& KAN & 3 & [32, 128, 256] & --- & $2\times 10^{-4}$ & 0.0 & SiLU \\
& Neural ODE & 2 & 256 & --- & $5\times 10^{-4}$ & 0.0 & SiLU \\
& Transformer & 4 & 256 & --- & $8\times 10^{-4}$ & 0.1 & SiLU \\
& Mamba & 3 & 256 & --- & $8\times 10^{-4}$ & 0.1 & SiLU \\
& DeepONet & 3+3$^\dagger$ & 256 & --- & $8\times 10^{-4}$ & 0.0 & Tanh \\
\midrule
\multirow{6}{*}{\shortstack[c]{Neural\\Operators}} 
& FNO & 4 & 64 & 16 & $8\times 10^{-4}$ & 0.0 & GELU \\
& MWT & 2 & 64 & --- & $8\times 10^{-4}$ & 0.0 & ReLU \\
& WNO & 4 & 64 & --- & $8\times 10^{-4}$ & 0.0 & GELU \\
& GINO & 4 & 64 & 16 & $1\times 10^{-3}$ & 0.0 & GELU \\
& LNO & 4 & 64 & 16 & $8\times 10^{-4}$ & 0.0 & GELU \\
& Tau-BNO & 2+2+2$^\ddagger$ & 64 & 16 & $8\times 10^{-4}$ & 0.0 & GELU \\
\bottomrule
\end{tabular}
\begin{tablenotes}
\small
\item $^\dagger$ 3 trunk layers + 3 branch layers
\item $^\ddagger$ 2 Function operator + 2 Query operator + 2 Directed Graph operator layers
\end{tablenotes}
\end{table}

Model checkpoints were saved based on validation loss, and we report test set performance using the best-performing checkpoint across all runs.

\subsubsection*{Biophysical Regimes.}

\begin{table}[htbp]
\centering
\caption{Description of biophysical regimes used to generate NTM simulations in Figures \ref{fig:r2-fig} and \ref{fig:seeding} (LH: left hemisphere, RH: right hemisphere).}
\label{tab:bio-regimes}
\small
\renewcommand{\arraystretch}{1.15}
\setlength{\tabcolsep}{4pt}
\begin{threeparttable}
\begin{tabular}{
    >{\centering\arraybackslash}m{0.11\linewidth}
    >{\centering\arraybackslash}m{0.20\linewidth}
    >{\centering\arraybackslash}m{0.08\linewidth}
    >{\centering\arraybackslash}m{0.08\linewidth}
    >{\centering\arraybackslash}m{0.03\linewidth}
    >{\centering\arraybackslash}m{0.03\linewidth}
    >{\centering\arraybackslash}m{0.03\linewidth}
    >{\centering\arraybackslash}m{0.22\linewidth}
}
\toprule
 & Seed Region(s)
 & $\lambda_f$\raisebox{+0.66ex}{\scalebox{0.8}{$(\times10^{-4})$}}
 & $\lambda_{\gamma}$\raisebox{+0.66ex}{\scalebox{0.8}{$(\times10^{-3})$}}
 & $\lambda_{\delta}$
 & $\lambda_{\epsilon}$
 & $\lambda_{\mu}$
 & Description \\
\midrule
Biophysical Regime 1&   Field CA1, LH&$100$&  $1.0$&  $10$&  $100$&  $2.2$&High production, low aggregation, high retrograde transport bias, high cellular uptake and release\\
\midrule
Biophysical Regime 2&   Field CA1, LH&$5$&  $1.0$&  $10$&  $10$&  $2.2$&Low production, low aggregation, low dibirectional transport, high cellular uptake and release\\
\midrule
Biophysical Regime 3&   Field CA1, LH&$0$&  $8.0$&  $100$&  $100$&  $3.2$&No production, high aggregation, high bidirectional transport, high cellular uptake and release\\
\midrule
Biophysical Regime 4&   Field CA1, LH&$0$&  $8.0$&  $100$&  $100$&  $0.2$&No production, high aggregation, high bidirectional transport, low cellular uptake and release\\
\midrule
Biophysical Regime 5& Caudoputamen,  Primary motor area; both RH& $1.8$& $7.2$& $81$& $18$& $1.0$&Low production, high aggregation, anterograde transport bias, low uptake and release\\
\midrule
Biophysical Regime 6& Medulla, Pons in LH; 
Amygdala, 
Auditory cortex, 
Hypothalamus, 
Medulla, 
Pons, 
Posterior parietal association area, 
Primary motor area, 
Restrosplenial area, 
Thalamic in RH.& $3.3$& $4.4$& $30$& $64$& $0.5$&Low production, intermediate aggregation, mild retrograde transport bias, low uptake and release\\
\midrule
Biophysical Regime 7& Ectorhinal area, 
Entorhinal area, 
Perirhinal area; Each in LH \& RH.& $94$& $7.5$& $39$& $74$& $2.3$&High production, high aggregation, mild retrograde transport bias, high uptake and release\\
\bottomrule
\end{tabular}
\end{threeparttable}
\end{table}

The biophysical regimes are presented in Table~\ref{tab:bio-regimes}.

\subsection*{Supplementary Proofs}
\subsubsection*{Differential Kernel Proof}\label{sec:diff-proof}

\begin{lemma}[Differential kernel]\label{lem:diff-kernel}
For any feature fields $Z$,
\begin{equation*}\label{eq:diff-kernel-identity}
  (\mathcal{K}_\ell^{\mathcal{D}}[Z])(x)=\nabla Z(x)\cdot b_\ell.
\end{equation*}
\end{lemma}

\textbf{proof.}
% \begin{proof}

We will prove that \(\big(\mathcal{K}_\ell^{\mathcal{D}}[Z]\big)(x) = \nabla Z(x)\,\cdot\, {b}_{\ell}\).

First, we define the convolution of the kernel \(K_\ell\) with the feature \(Z\) as
\[
(\mathrm{Conv}_{K_{\ell}} Z)(x) = \sum_i (K_{\ell})_i(x,\xi) Z(\xi).
\]
Next, applying a Taylor expansion around \(x\), we express \(Z(\xi)\) as
\[
Z(\xi) = Z(x) + \nabla Z(x)(\xi - x) + \mathcal{O}(h).
\]
Substituting this into the convolution expression, we get:
\[
(\mathrm{Conv}_{K_{\ell}} Z)(x) \approx \sum_i (K_\ell)_i(x,\xi) \left[ Z(x) + \nabla Z(x)(\xi - x) \right].
\]
By separating the terms, we obtain:
\begin{equation*}
(\mathrm{Conv}_{K_\ell} Z)(x) 
= \left[ \sum_i (K_\ell)_i(x,\xi) \right] Z(x) + \left[ \sum_i (K_\ell)_i(x,\xi)(\xi - x) \right] \nabla Z(x)
= Z(x)\cdot c_\ell + \nabla Z(x) \cdot b_\ell,
\end{equation*}
where 
\[
c_\ell = \sum_i (K_\ell)_i(x,\xi).
\]
We note that by subtracting the mean value \(\overline{K}_\ell = \sum_i (K_\ell)_i(x,\xi)\) from the kernel, we can rewrite \(c_\ell\) as follows:
\[
c_\ell = \sum_i (K_\ell)_i(x,\xi) - \overline{K}_\ell(x,\xi) = 0.
\]
For the expression of \(b_\ell\), we can define the kernel moments as:
\[
b_\ell = \sum_i (K_\ell)_i(x,\xi)(\xi - x),
\]
where \(b_\ell\) captures the learned moments of the kernel.
Therefore, we conclude that the differential kernel converges to the first-order derivative operator, as stated in the theorem.

% \end{proof}

\subsection*{Supplementary Results}\label{sec:extra_results}
\subsubsection*{Trajectory Prediction Across Diverse Physiological Regimes}\label{sec:Another_Trajectory_Prediction}

\begin{figure}[htbp]
\centering
\includegraphics[width=0.95\textwidth]{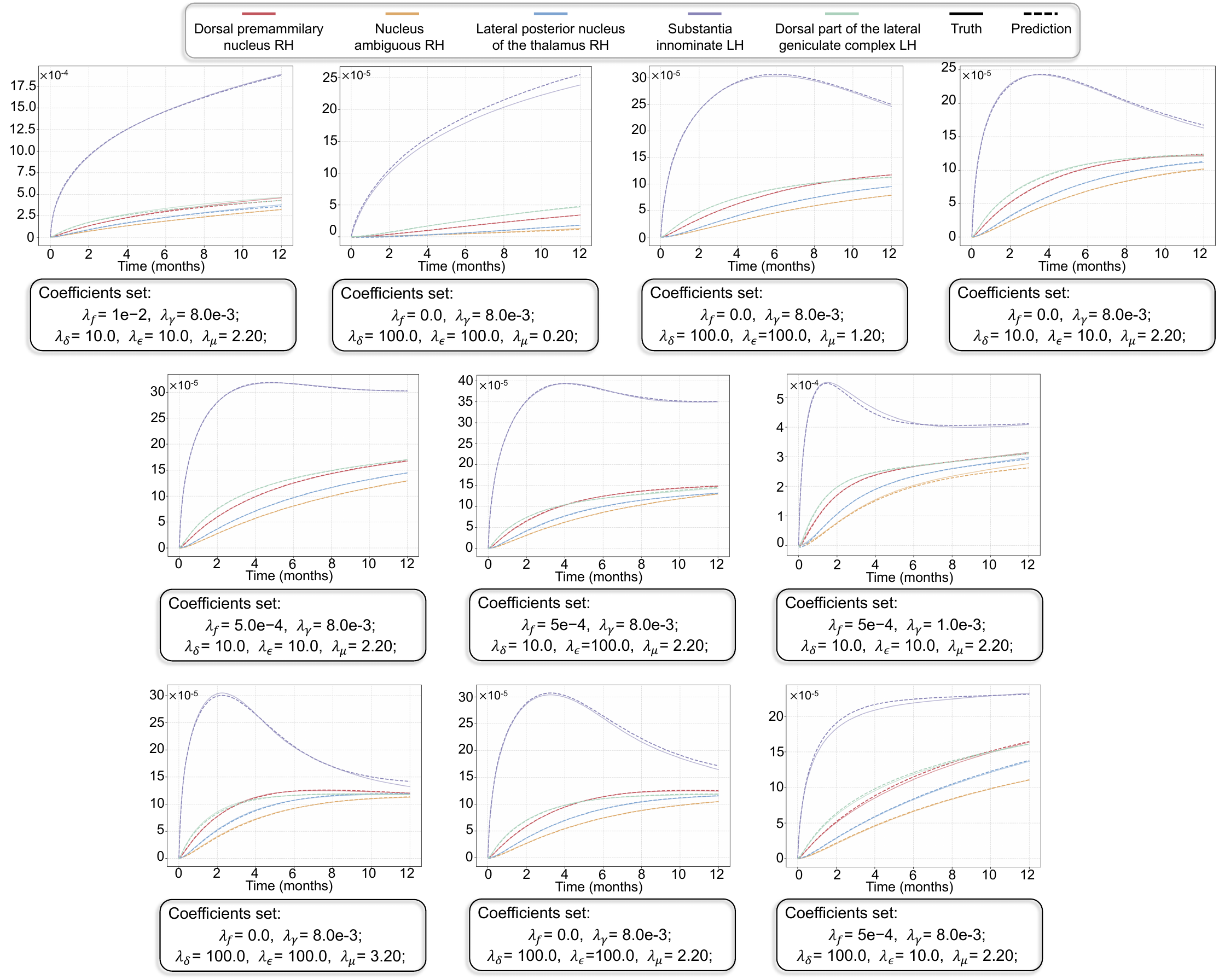}
\caption{\textbf{Trajectory predictions across ten diverse physiological regimes.} 
Dashed lines represent Tau-BNO predictions; solid lines indicate ground truth trajectories from high-fidelity PDE solutions. 
The model demonstrates robust performance across parameter conditions for five representative brain regions, accurately capturing both transient dynamics and steady-state behavior. Each condition represents distinct combinations of production rate ($\lambda_f$), aggregation rate ($\lambda_\gamma$), transport parameters ($\lambda_{\delta}$, $\lambda_{\epsilon}$), and uptake release rate ($\lambda_\mu$).}
\label{fig:Results_Extra_TenExample_Trajectory}
\end{figure}

To evaluate the generalization capacity of Tau-BNO, we tested its performance on ten distinct parameter configurations across five representative brain regions (as shown in Figure~\ref{fig:Results_Extra_TenExample_Trajectory}). These configurations comprehensively sample the physiological parameter space, varying tau production rates, aggregation kinetics, and directional transport parameters that govern pathological tau propagation.
This analysis demonstrates that (i) different parameter combinations produce distinct pathological trajectories matching diverse tauopathy phenotypes, and (ii) neural operators can efficiently replace traditional PDE solvers for exploring high-dimensional parameter spaces in precision medicine.
Figure~\ref{fig:Results_Extra_TenExample_Trajectory} shows that Tau-BNO achieves excellent trajectory reconstruction across all regimes, with a mean relative error of 1.6\%. 
By combining solver-level accuracy with deep learning efficiency, Tau-BNO opens the door to large-scale system parameter exploration and mechanistic investigation of tau propagation dynamics, tasks long hindered by the prohibitive cost of repeated PDE solves.

\clearpage

\printbibliography

%\appendix
%\section{Appendix}
%Appendices (optional).

\end{document}